\documentclass[reprint,twocolumn]{revtex4}
\usepackage{amsfonts}
\usepackage{amsmath}
\usepackage{amssymb}
\usepackage{charter}
\usepackage{subfigure}
\usepackage{graphicx}
\usepackage{float}

\usepackage[T1]{fontenc}
\usepackage[utf8]{inputenc}

\begin{document}

\title{Impact of the Unruh effect on the estimation precision of Gaussian
channel parameters}
\author{Shoukang Chang$^{1}$}
\author{Yawen Tang$^{1}$}
\author{Wei Ye$^{2}$}
\author{Shao-Ming Fei$^{1,3}$}
\author{Zunlue Zhu$^{1}$}
\thanks{Contact author: zl-zhu@htu.cn}
\author{Xingdong Zhao$^{1}$}
\thanks{Contact author: phyzhxd@gmail.com}
\affiliation{$^{{\small 1}}$\textit{School of Physics, Henan Normal University, Xinxiang
453007, China}\\
$^{{\small 2}}$\textit{School of Information Engineering, Nanchang Hangkong
University, Nanchang 330063, China}\\
$^{{\small 3}}$\textit{School of Mathematical Sciences, Capital Normal
University, Beijing 100048, China}}

\begin{abstract}
Gaussian quantum channels constitute a pivotal physical framework for
characterizing the dynamics of Gaussian quantum states. Extensive scholarly
attention has been devoted to the estimation of parameters associated with
Gaussian channels. However, while previous research has predominantly
focused on parameter estimation within inertial frames, the non-inertial
scenario, particularly in the context of the Unruh effect, remains largely
unexplored. In this paper, we analyze the impact of the Unruh effect on the
estimation precision of Gaussian channel parameters, with a specific focus
on thermal attenuator and thermal amplifier channels. Our findings reveal
that the Unruh effect significantly degrades the precision of
single-parameter estimation for Gaussian channel parameters when employing
both the input coherent state and squeezed vacuum state. For the
two-parameter estimation, we further demonstrate that the quantum Cram\'{e}%
r-Rao bound serves as an asymptotically achievable precision limit.
Consistent with the single-parameter case, the Unruh effect exerts a
detrimental impact on the precision of two-parameter estimation. Notably,
heterodyne measurement is near-optimal for both single- and two-parameter
estimation in the limit of high acceleration or large thermal mean numbers.
These results provide crucial theoretical insights and practical guidance
for advancing quantum parameter estimation in relativistic context.
\end{abstract}

\maketitle

\section{Introduction}

Quantum parameter estimation aims to harness distinctive quantum phenomena,
including quantum entanglement, coherence, and squeezing, to transcend the
precision limits imposed by classical methodologies \cite{1,2,3,4}. The
theoretical foundation of quantum parameter estimation is fundamentally
grounded in the quantum Cram\'{e}r-Rao bound (QCRB), a pivotal framework
that delineates the ultimate precision limits attainable in quantum
metrological processes \cite{5,6,7,8}. The inverse of the QCRB, referred to
as the quantum Fisher information, acts as a pivotal measure that assesses
the responsiveness of a quantum system to minor parameter variations \cite%
{9,10}. In addition to its fundamental importance in quantum parameter
estimation, the quantum Fisher information has emerged as a multifaceted and
robust analytical instrument. It is now widely utilized in advanced fields
like distributed quantum sensing \cite{11,12}, quantum speed limit \cite%
{13,14,15}, and optical imaging \cite{16,17}.

The expansive range of potential applications in quantum parameter
estimation has significantly driven the development of theoretical
methodologies for assessing the QCRB \cite{18}. In this context, an
analytical framework has been established to derive the QCRB specifically
tailored for Gaussian quantum systems \cite{19,20,21,22,23}. This approach
has been widely applied in various estimation scenarios, including but not
limited to Gaussian channel parameters \cite{24,25,26,27,28,29}, damping
rates \cite{30,31}, temperature \cite{32,33,34}, and optomechanical
parameters \cite{18,35} within Gaussian quantum systems. Notably, the
estimation of Gaussian channel parameters has recently garnered substantial
academic attention. A landmark contribution was made by Monras and Paris,
who provided the first comprehensive solution to the loss parameter
estimation problem using single-mode pure Gaussian states, where the QCRB
serves as a crucial metric for estimation precision \cite{24}. Expanding on
this foundation, a diverse array of estimation strategies for Gaussian
channel parameters has been developed, including the use of non-Gaussian
probes \cite{36,37}, the leveraging of quantum coherence \cite{28}, the
employment of ring resonators \cite{38,39}, and the simultaneous estimation
of multiple loss parameters \cite{29}. However, these research endeavors
have predominantly focused on the estimation of Gaussian channel parameters
within an inertial frame. As a result, the estimation of Gaussian channel
parameters in a non-inertial frame remains an unresolved and intriguing
challenge in the field.

On the other hand, relativistic quantum information, which explores quantum
information processes and concepts within a relativistic context, has
emerged as a vibrant and rapidly evolving field of research, driven by both
theoretical and experimental imperatives \cite{40,41}. This framework is
essential for comprehending quantum phenomena, as realistic quantum systems
inherently operate in non-inertial environments \cite{42}. A striking
example of this is the Unruh effect, which predicts that an accelerating
observer in a vacuum will perceive a thermal bath of particles, whereas an
inertial observer would detect nothing \cite{43,44,45}. This phenomenon
highlights the profound interplay between quantum field theory and
relativity, revealing how the perception of quantum states can drastically
differ depending on the observer's motion \cite{43,44,45}. From the
perspective of relativistic frameworks, the Unruh effect inevitably leads to
the degradation of quantum information, including the loss of quantum
entanglement \cite{46,47}, quantum steering \cite{42}, quantum coherence 
\cite{48,49}, quantum communication \cite{50,51,52} and quantum parameter
estimation precision \cite{53,54,55,56,57}. However, as highlighted in Ref. 
\cite{58}, the Unruh effect can also generate net quantum entanglement,
contingent on the choice of the inertial state. The corresponding Unruh
effect cannot be simplistically characterized as a conventional noisy
channel, as it gives rise to counterintuitive and nuanced phenomena that
challenge traditional assumptions \cite{58}. Consequently, how the Unruh
effect influences the estimation precision of the QCRB is a compelling and
underexplored dimension in quantum parameter estimation \cite{54}.

In this paper, we explore the impact of the Unruh effect on the precision of
parameter estimation in Gaussian channels, specifically examining thermal
attenuator and thermal amplifier channels. Our results reveal that the Unruh
effect substantially diminishes the estimation precision of both the
attenuation parameter and the amplifier parameter when utilizing coherent
states and squeezed vacuum states as input states. Specifically, we observe
that the quantum Fisher information decreases monotonically as the
acceleration parameter increases. Furthermore, we find that heterodyne
measurement outperforms homodyne measurement and approaches optimality in
the limits of high acceleration parameters or high thermal mean numbers.%
\textbf{\ }Furthermore, we investigate a two-parameter estimation problem in
two specific cases: i) the joint estimation of the attenuation parameter and
the thermal mean number of the thermal attenuator channel; and ii) the joint
estimation of the amplifier parameter and the thermal mean number of the
thermal amplifier channel. Our analysis reveals that the symmetric
logarithmic derivative operators \cite{5,59} do not commute, and the mean
Uhlmann curvature matrix \cite{59,60,61} is a zero matrix, indicating that
the QCRB represents an asymptotically achievable precision limit. Consistent
with the single-parameter scenario, the Unruh effect adversely affects the
precision of two-parameter estimation. Moreover, we find that heterodyne
measurement is nearly optimal under conditions of high acceleration or high
thermal mean numbers. Overall, our findings highlight the detrimental role
of the Unruh effect in both single- and two-parameter estimation scenarios
across different Gaussian channel configurations, providing valuable
insights into the interplay between relativistic effects and quantum
parameter estimation.

The remainder of this paper is arranged as follows. In Sec. II, we review
the known results of Gaussian quantum parameter estimation theory. In Sec.
III, we explore the quantum estimation of Gaussian channel parameters in the
context of Unruh effect. Finally, our main conclusions are drawn in the last
section.

\section{Gaussian quantum parameter estimation}

In this section, we commence with a systematic introduction to Gaussian
quantum systems, involving both Gaussian quantum states and Gaussian quantum
channels \cite{62,63,64}. Subsequently, we elaborate on the fundamental
principles of quantum parameter estimation, which provides the theoretical
framework for determining the ultimate precision bounds associated with
Gaussian channel parameters.

\subsection{Gaussian quantum systems}

Let $\hat{r}=(\hat{x}_{1},\hat{p}_{1},...,\hat{x}_{m},\hat{p}_{m})^{T}$ be a
vector of quadrature operators, which can be used for characterizing a $m$%
-mode bosonic Gaussian quantum systems. The corresponding canonical
commutation relation is defined as%
\begin{equation}
\lbrack \hat{r},\hat{r}^{T}]=i\Omega ,  \label{1}
\end{equation}%
where $\Omega =\oplus _{j=1}^{m}\Omega _{1}$ is the symplectic matrix with $%
\Omega _{1}=\left( 
\begin{array}{cc}
0 & 1 \\ 
-1 & 0%
\end{array}%
\right) .$ For the sake of discussion and analysis, we employ natural units (%
h{\hskip-.2em}\llap{\protect\rule[1.1ex]{.325em}{.1ex}}{\hskip.2em}%
=$k_{B}$=$1$). The Gaussian quantum states in phase space can be
equivalently characterized by the characteristic function%
\begin{equation}
\chi _{G}(r)=\exp \left[ -\frac{1}{4}\tilde{r}^{T}\sigma \tilde{r}+i\tilde{r}%
^{T}d\right] ,  \label{2}
\end{equation}%
where $\tilde{r}=\Omega r$ with $r=(x_{1},p_{1},...,x_{m},p_{m})^{T}$ is a
vector of 2$m$ real coordinates in phase-space, $d=$Tr($\hat{\rho}\hat{r}$)
is the first moment, and $\sigma $ is the covariance matrix%
\begin{equation}
\sigma =\text{Tr}\left[ \hat{\rho}\left\{ (\hat{r}-d),(\hat{r}%
-d)^{T}\right\} \right] \text{,}  \label{3}
\end{equation}%
where the notation $\left\{ \cdot ,\cdot \right\} $ and Tr denote the
anticommutator and the trace of an operator in Hilbert space, respectively.
It is worth noting that the complete characterization of any Gaussian
quantum state can be fully determined by its first moment $d$ and covariance
matrix $\sigma .$

When any Gaussian quantum state is input into a Gaussian quantum channel $%
\Lambda $, the corresponding transformation process can be completely
described by two 2$m\times $ 2$m$ real symmetric matrices, the scaling
matrix $X$ and the noise matrix $Y$, i.e.,

\begin{eqnarray}
d &\mapsto &Xd,  \notag \\
\sigma &\mapsto &X\sigma X^{T}+Y,  \label{4}
\end{eqnarray}%
which satisfy the complete positivity condition%
\begin{equation}
Y+iX\Omega X^{T}\geq i\Omega .  \label{5}
\end{equation}%
The specific forms of the matrices $X$ and $Y$ are directly determined by
the type of quantum channel, such as for the single-mode thermal attenuator
channel, 
\begin{eqnarray}
X_{att} &=&\cos \vartheta I_{2},  \notag \\
Y_{att} &=&(2\bar{N}_{\text{att}}+1)\sin ^{2}\vartheta I_{2},  \label{6}
\end{eqnarray}%
where $I_{2}$ is the $2\times 2$ identity matrix, $\vartheta \in \lbrack
0,2\pi ]$ represents the attenuation parameter, and $\bar{N}_{\text{att}%
}\geq 0$ denotes the thermal mean number of the thermal attenuator channel,
which is the fundamental characteristic parameter defining the thermal
attenuator channel. If $\bar{N}_{\text{att}}=0,$ such a channel is
occasionally termed as a "quantum limited attenuator", which characterizes
the effect on the Gaussian quantum system of a zero-temperature bath \cite%
{64}. Another important quantum channel is the thermal amplifier channel,
which can also be described by%
\begin{eqnarray}
X_{\text{amp}} &=&\cosh \epsilon I_{2},  \notag \\
Y_{\text{amp}} &=&(2\bar{N}_{\text{amp}}+1)\sinh ^{2}\epsilon I_{2},
\label{7}
\end{eqnarray}%
where $\epsilon $ and $\bar{N}_{\text{amp}}$ are respectively the amplifier
parameter and the thermal mean number of the thermal amplifier channel. If $%
\bar{N}_{\text{amp}}=0,$ the channel is referred to as a "quantum limited
amplifier", which amplifies the optical signal of the input states \cite{64}.

\subsection{Quantum parameter estimation}

Let us consider a quantum state $\hat{\rho}_{\theta }$ encoded into an
unknown parameter $\theta $ to be estimated. The corresponding estimation
precision, quantified by the mean square error, is bounded by the Cram\'{e}%
r-Rao inequality

\begin{equation}
Var(\theta )\geq \frac{1}{\tilde{F}_{\theta }}\geq \frac{1}{F_{\theta }},
\label{8}
\end{equation}%
where $\tilde{F}_{\theta }$\ and $F_{\theta }$\ denote the classical and
quantum Fisher information (CFI and QFI), respectively. For more
comprehensive explanations and technical details on the Cram\'{e}r-Rao
inequality, we refer readers to the Refs. \cite{5,6,7,8,59,61}.

When multiple unknown parameters $\tilde{\theta}$=$(\theta _{1},...,\theta
_{d})^{T}$\ are estimated simultaneously, the precision is governed by the
mean square error matrix $\Sigma (\tilde{\theta}),$\ which satisfies the
following matrix inequality%
\begin{equation}
\Sigma (\tilde{\theta})\geq \tilde{J}^{-1}\geq J^{-1},  \label{9}
\end{equation}%
where $\tilde{J}$\ is the CFI matrix, whose elements are%
\begin{equation}
\tilde{J}_{uv}=\int dxp(x|\theta )\left( \frac{\partial \ln p(x|\theta )}{%
\partial \theta _{u}}\right) \left( \frac{\partial \ln p(x|\theta )}{%
\partial \theta _{v}}\right) ,  \label{10}
\end{equation}%
with $p(x|\theta )$\ being the conditional probability distribution.
Correspondingly, $J$ represents the quantum Fisher information matrix with
elements%
\begin{equation}
J_{uv}=\frac{1}{2}\text{Tr}\left[ \hat{\rho}_{\tilde{\theta}}(\hat{L}_{u}%
\hat{L}_{v}+\hat{L}_{v}\hat{L}_{u})\right] \text{,}  \label{11}
\end{equation}%
where $\hat{L}_{u}$ are the symmetric logarithmic derivative operators
determined by the Lyapunov equation $\left. \partial \hat{\rho}_{\tilde{%
\theta}}\right/ \partial \theta _{u}=\left. (\hat{L}_{u}\hat{\rho}_{\tilde{%
\theta}}+\hat{\rho}_{\tilde{\theta}}\hat{L}_{u})\right/ 2.$ If the quantum
state $\hat{\rho}_{\tilde{\theta}}$\ is restricted to a Gaussian quantum
state characterized by its first moment $d_{\tilde{\theta}}$\ and covariance
matrix $\sigma _{\tilde{\theta}}$, and Gaussian measurements are
implemented, the elements of classical Fisher information matrix can be
expressed as \cite{65} 
\begin{eqnarray}
\tilde{J}_{uv} &=&\left( \frac{\partial d_{\tilde{\theta}}}{\partial \theta
_{u}}\right) ^{T}\Sigma _{\tilde{\theta}}^{-1}\left( \frac{\partial d_{%
\tilde{\theta}}}{\partial \theta _{v}}\right)  \notag \\
&&+\frac{1}{2}\text{tr}\left[ \Sigma _{\tilde{\theta}}^{-1}\left( \frac{%
\partial \Sigma _{\tilde{\theta}}}{\partial \theta _{u}}\right) \Sigma _{%
\tilde{\theta}}^{-1}\left( \frac{\partial \Sigma _{\tilde{\theta}}}{\partial
\theta _{v}}\right) \right] ,  \label{12}
\end{eqnarray}%
where tr is the trace of matrix and $\Sigma _{\tilde{\theta}}=\left. (\sigma
_{\tilde{\theta}}+\sigma _{m})\right/ 2.$\ In a single-mode scenario, $%
\sigma _{m}=$diag$\left( z,1/z\right) $\ defines the specific measurement
scheme where $z=1$\ and\ $z\rightarrow 0$\ correspond to heterodyne and
homodyne measurements, respectively.\textbf{\ }Likewise, for a Gaussian
quantum state, the symmetric logarithmic derivative operators $\hat{L}_{u}$
manifest as Hermitian quadratic operators \cite{19,20,21,22,23,64} 
\begin{equation}
\hat{L}_{u}=L_{u}^{(0)}\text{\^{I}}+\left( L_{u}^{(1)}\right) ^{T}\hat{r}+%
\hat{r}^{T}L_{u}^{(2)}\hat{r},  \label{13}
\end{equation}%
where%
\begin{eqnarray}
L_{u}^{(0)} &=&-\frac{1}{2}\text{tr}\left[ \sigma _{\tilde{\theta}%
}L_{u}^{(2)}\right] -\left( L_{u}^{(1)}\right) ^{T}d_{\tilde{\theta}}-d_{%
\tilde{\theta}}^{T}L_{u}^{(2)}d_{\tilde{\theta}},  \notag \\
L_{u}^{(1)} &=&2\sigma _{\tilde{\theta}}^{-1}\frac{\partial d_{\tilde{\theta}%
}}{\partial \theta _{u}}-2L_{u}^{(2)}d_{\tilde{\theta}},  \notag \\
\text{vec}[L_{u}^{(2)}] &=&(\sigma _{\tilde{\theta}}\otimes \sigma _{\tilde{%
\theta}}-\Omega \otimes \Omega )^{-1}\text{vec}\left[ \frac{\partial \sigma
_{\tilde{\theta}}}{\partial \theta _{u}}\right] \text{,}  \label{14}
\end{eqnarray}%
with vec$[\cdot ]$ being the vectorization of a matrix $A$, i.e., the column
vector constructed from columns of the following matrix%
\begin{equation}
A=\left( 
\begin{array}{cc}
a_{11} & a_{12} \\ 
a_{21} & a_{22}%
\end{array}%
\right) \longrightarrow \text{vec}[A]=\left( 
\begin{array}{c}
a_{11} \\ 
a_{21} \\ 
a_{12} \\ 
a_{22}%
\end{array}%
\right) .  \label{15}
\end{equation}%
The corresponding quantum Fisher information matrix can be also expressed in
terms of the first moment and the covariance matrix \cite{19,20,21,22,23} 
\begin{eqnarray}
J_{uv} &=&\frac{1}{2}\text{vec}\left[ \frac{\partial \sigma _{\tilde{\theta}}%
}{\partial \theta _{u}}\right] ^{T}(\sigma _{\tilde{\theta}}\otimes \sigma _{%
\tilde{\theta}}-\Omega \otimes \Omega )^{-1}\text{vec}\left[ \frac{\partial
\sigma _{\tilde{\theta}}}{\partial \theta _{v}}\right]  \notag \\
&&+2\left( \frac{\partial d_{\tilde{\theta}}}{\partial \theta _{u}}\right)
^{T}\sigma _{\tilde{\theta}}^{-1}\frac{\partial d_{\tilde{\theta}}}{\partial
\theta _{v}}.  \label{16}
\end{eqnarray}%
Since scalar bounds are generally more convenient to handle, we introduce a
real, positive, $d\times d$\ weight matrix $W$\ to derive the following
scalar inequality%
\begin{equation}
\text{tr[}W\Sigma (\tilde{\theta})\text{]}\geq C\geq Q,  \label{17}
\end{equation}%
where $C\equiv $tr[$W\tilde{J}^{-1}$] and $Q\equiv $tr[$WJ^{-1}$] are the
Cram\'{e}r-Rao bound (CRB) and QCRB, respectively. In this paper, we take $%
W=I_{d}$\ to be the identity matrix, thereby ensuring that the inequality
bounds tr[$W\Sigma (\tilde{\theta})$] $\geq C\geq Q$, which represents the
sum of the variances of each parameter estimation \cite{65}. This choice is
natural in a generic setting, where the estimation of each parameter is
considered equally important a priori \cite{65}. It is noteworthy that while
the CRB can be asymptotically saturated by an efficient estimator, the QCRB
is generally not tight in multi-parameter estimation, although it is always
tight for the single-parameter case \cite{59,61}. The QCRB becomes tight if
and only if the SLD operators commute, i.e., $[\hat{L}_{u},\hat{L}_{v}]=0$\
for all $u$\ and $v$\ \cite{59,61}. In such cases, a common basis of
eigenstates exists, forming an optimal measurement basis to saturate the
QCRB via single-copy measurements. In addition, there exist some special
quantum states that satisfy $[\hat{L}_{u},\hat{L}_{v}]\neq 0$ and the mean
Uhlmann curvature matrix with elements 
\begin{equation}
D_{uv}=-\frac{i}{2}\text{Tr}\left[ \hat{\rho}_{\tilde{\theta}}[\hat{L}_{u},%
\hat{L}_{v}]\right] ,  \label{18}
\end{equation}%
is a zero matrix \cite{59,61,66,67,68}. The corresponding QCRB is
asymptotically tight and can be saturated asymptotically through the
implementation of collective measurements \cite{59,61,66,67,68,69}.
Furthermore, the QCRB and the Holevo Cram\'{e}r-Rao bound are numerically
the same in this case \cite{59,61}. In particular, for a Gaussian quantum
state $\hat{\rho}_{\tilde{\theta}},$ the commutation relation $[\hat{L}_{u},%
\hat{L}_{v}]$ and the mean Uhlmann curvature matrix elements $D_{uv}$
respectively have the following specific form \cite{19,20,21,22,23} 
\begin{eqnarray}
\lbrack \hat{L}_{u},\hat{L}_{v}] &=&i[\left( L_{u}^{(1)}\right) ^{T}\Omega
L_{v}^{(1)}+2\left( L_{u}^{(1)}\right) ^{T}\Omega L_{v}^{(2)}\hat{r}  \notag
\\
&&+2\hat{r}^{T}L_{u}^{(2)}\Omega L_{v}^{(1)}+2\hat{r}^{T}L_{u}^{(2)}\Omega
L_{v}^{(2)}\hat{r}  \notag \\
&&-2\hat{r}^{T}L_{v}^{(2)}\Omega L_{u}^{(2)}\hat{r}],  \notag \\
D_{uv} &=&\text{vec}[L_{u}^{(2)}]^{T}\left( \sigma _{\tilde{\theta}}\otimes
\Omega \right) \text{vec}[L_{v}^{(2)}]  \notag \\
&&+2\left( \frac{\partial d_{\tilde{\theta}}}{\partial \theta _{u}}\right)
^{T}\sigma _{\tilde{\theta}}^{-1}\Omega \sigma _{\tilde{\theta}}^{-1}\frac{%
\partial d_{\tilde{\theta}}}{\partial \theta _{v}}.  \label{19}
\end{eqnarray}%
In fact, these two conditions are equivalent for parameters encoded solely
in the first moments; however, the full commutativity of symmetric
logarithmic derivative operators represents a more stringent requirement
when the parameter dependence extends to the covariance matrix \cite{22}.

\section{Quantum estimation of Gaussian channel parameters in the context of
Unruh effect}

In this section, we analyze the impact of the Unruh effect on the estimation
of Gaussian channel parameters, with special attention given to thermal
attenuator and thermal amplifier channels. Prior to introducing the
parameter estimation of these specific channels, we first provide a concise
overview of known results regarding the quantum parameter estimation in the
context of Unruh effect \cite{53,54,55,56,57}. 
\begin{figure}[tbp]
\label{Fig1} \centering\includegraphics[width=0.8\columnwidth]{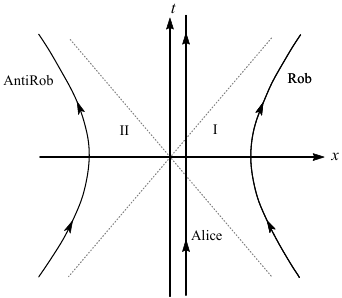}%
\newline
\caption{{}(Color online) Schematic diagram of the field mode evolution. The
diagram contrasts Alice's inertial perspective with the bipartite coupling
between Rob (Region I) and Anti-Rob (Region II) induced by the Unruh effect
across the causal horizon \protect\cite{55,58}.}
\end{figure}
\ \ \ \ \ \ \ \ \ Specifically, we consider a scenario involving an inertial
observer, Alice, and a uniformly accelerated observer, Rob, as illustrated
in the schematic diagram in Fig. 1. From the global perspective of an
inertial observer, the quantum state $\hat{\rho}_{0}$\ is assumed to undergo
an ideal unitary evolution via a Bogoliubov transformation. However, for the
non-inertial observer Rob, the perception of this state is fundamentally
altered by the existence of a causal horizon. As illustrated in Fig. 1, the
Unruh effect physically manifests as the spontaneous creation of entangled
particle pairs across this horizon. In this configuration, Alice prepares
the initial Gaussian state within the accessible Rindler wedge (Region I).
However, the physical modes in this region are intrinsically coupled to a
complementary set of modes residing in the causally disconnected wedge
(Region II), where a virtual observer, Anti-Rob, is situated. From a
physical standpoint, the Minkowski vacuum appears to Rob as a thermal bath
because his local observations are entangled with the modes confined to
Region II. Since Rob cannot perform measurements beyond the horizon, the
information shared with Anti-Rob is effectively lost to the environment. The
relationship between Alice's operators and the Rindler operators $\hat{b}_{%
\text{I}}$\ (associated with Rob) and $\hat{b}_{\text{II}}$\ (associated
with Anti-Rob) is governed by a Bogoliubov transformation \cite{53,54,55} 
\begin{eqnarray}
\hat{a}_{\text{I}} &=&\hat{b}_{\text{I}}\cosh r-\hat{b}_{\text{II}}^{\dagger
}\sinh r,  \notag \\
\hat{a}_{\text{II}} &=&\hat{b}_{\text{II}}\cosh r-\hat{b}_{\text{I}%
}^{\dagger }\sinh r,  \label{20}
\end{eqnarray}%
where $r$ is the "acceleration parameter" and maintains a direct
proportionality with the Unruh temperature $\cosh ^{-2}r=1-e^{-\omega /T}.$
By tracing over the inaccessible modes in Region II, the initial state $\hat{%
\rho}_{0}$\ transforms into a noisy output state $\hat{\rho}_{r}$\ from
Rob's viewpoint \cite{53} 
\begin{equation}
\hat{\rho}_{r}=\text{Tr}_{\text{II}}[\hat{U}(r)(\hat{\rho}_{0}\otimes
\left\vert 0\right\rangle \left\langle 0\right\vert )\hat{U}^{\dagger }(r)],
\label{21}
\end{equation}%
where $\hat{U}(r)=e^{r(\hat{b}_{\text{I}}^{\dagger }\hat{b}_{\text{II}%
}^{\dagger }-\hat{b}_{\text{I}}\hat{b}_{\text{II}})}$\ represents the
two-mode squeezing operator encoding the correlation across the horizon.
This non-unitary transition is mathematically equivalent to a bosonic
amplification channel. When subjected to the bosonic amplification channel
induced by the Unruh effect, a Gaussian quantum state undergoes a
transformation, evolving into another Gaussian quantum state, which is fully
characterized by its first moment and covariance matrix \cite{53} 
\begin{eqnarray}
d_{r} &=&X_{r}d_{0},  \notag \\
\sigma _{r} &=&X_{r}\sigma _{0}X_{r}^{T}+Y_{r},  \label{22}
\end{eqnarray}%
with $X_{r}=\cosh rI_{2}$ and $Y_{r}=\sinh ^{2}rI_{2}.$ It is worth noting
that the theoretical framework of this study employs the single-mode
approximation to describe the relationship between Minkowski and Rindler
modes. In this approximation, a specific Minkowski frequency mode is assumed
to be related to a single Rindler mode through the Bogoliubov
transformation. However, it is essential to consider the findings of Bruschi
et al., who demonstrated that the mapping between these frames is
fundamentally multimode \cite{70}. Despite the inherent complexity of the
multimode case, the single-mode treatment provides a crucial analytical
benchmark for evaluating the precision limits of Gaussian channels. This
regime remains physically relevant for scenarios that satisfy the peaking
constraints set by an appropriate Fourier transform \cite{70}.

\subsection{Thermal attenuator channel}

Here we focus on two quantum states as input states: the coherent state and
the squeezed vacuum state, both of which are completely characterized by
their respective first moments and covariance matrices%
\begin{eqnarray}
d_{\alpha } &=&\sqrt{2}(\text{Re(}\alpha \text{), Im(}\alpha \text{)})^{T}, 
\notag \\
\sigma _{\alpha } &=&I_{2},  \notag \\
d_{s} &=&(\text{0, 0})^{T},  \notag \\
\sigma _{s} &=&\left( 
\begin{array}{cc}
e^{-2s} & 0 \\ 
0 & e^{2s}%
\end{array}%
\right) ,  \label{23}
\end{eqnarray}%
where $\alpha $ is a complex number which corresponds to the complex
amplitude of the coherent state and $s$ is the squeezing parameter of the
squeezed vacuum state. Then, these two input states respectively undergo a
thermal attenuator channel to generate two probe states, which will depend
on the attenuation parameter $\vartheta $ and the thermal mean number $\bar{N%
}_{\text{att}}$ of the thermal attenuator channel$.$ Assuming that Alice
possesses the capability to prepare these two probe states within an
inertial frame, the output quantum states accessible to Rob are
fundamentally determined through the action of the bosonic amplification
channel, which is intrinsically induced by the Unruh effect, as explicitly
described by Eq. (\ref{21}). Within the framework of Gaussian formalism, the
complete evolution process is entirely governed by the first moment and
covariance matrix 
\begin{figure}[tbp]
\label{Fig2} \centering\includegraphics[width=0.8\columnwidth]{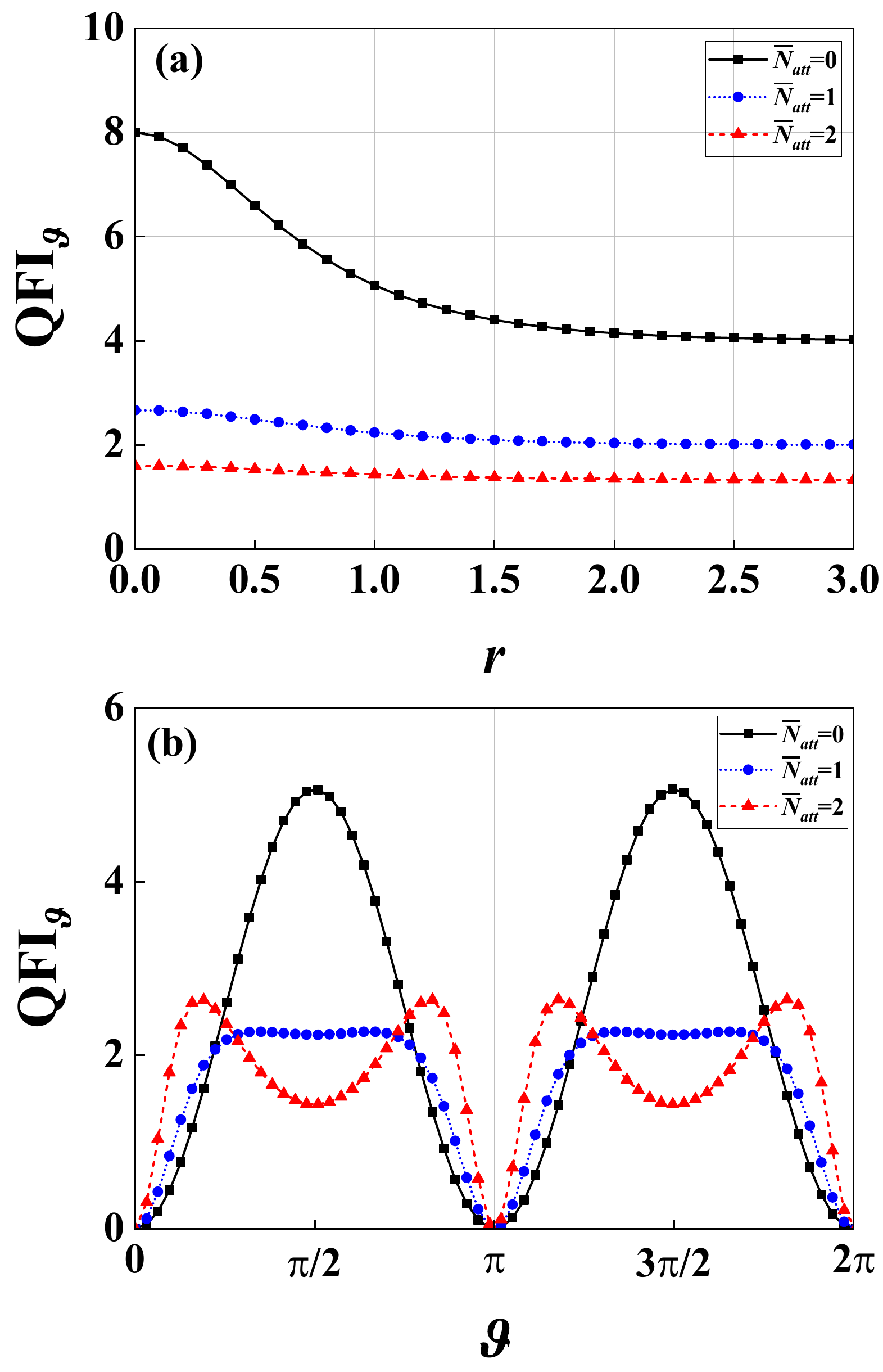}%
\newline
\caption{{}(Color online) The QFI with respect to the attenuation parameter $%
\protect\vartheta $ for the input coherent state is analyzed as a function
of (a) the acceleration parameter $r$\ with $\protect\vartheta =\protect\pi %
/2$ and\ $\left\vert \protect\alpha \right\vert ^{2}=2,$ and of (b) the
attenuation parameter $\protect\vartheta $\ with $r=1$ and\ $\left\vert 
\protect\alpha \right\vert ^{2}=2.$}
\end{figure}
\begin{figure}[tbp]
\label{Fig3} \centering\includegraphics[width=0.8\columnwidth]{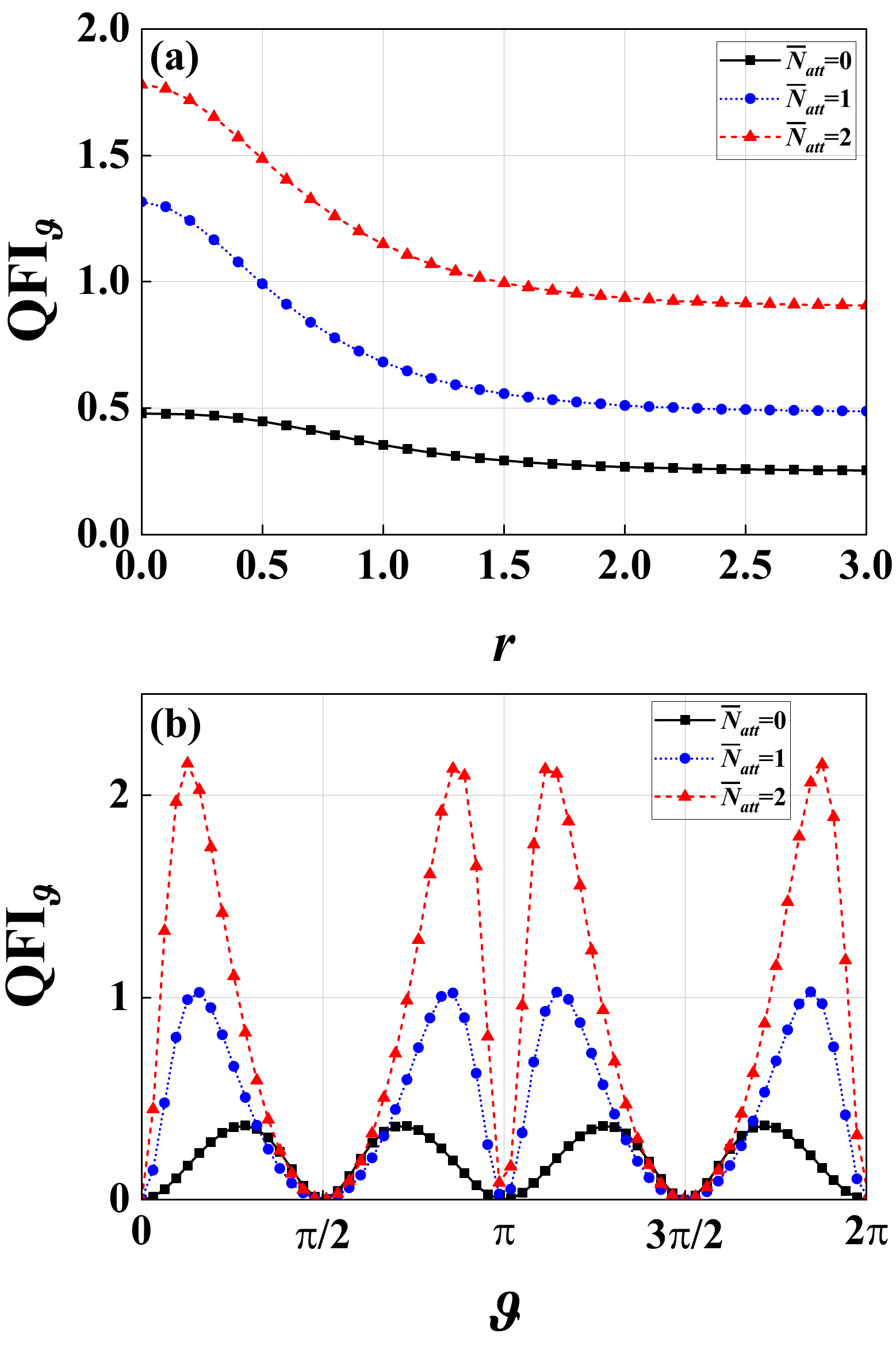}%
\newline
\caption{{}(Color online) The QFI with respect to the attenuation parameter $%
\protect\vartheta $ for the input squeezed vacuum state is analyzed as a
function of (a) the acceleration parameter $r$\ with $\protect\vartheta =%
\protect\pi /4$ and\ $s=0.5,$ and of (b) the attenuation parameter $\protect%
\vartheta $\ with $r=1$ and\ $s=0.5.$}
\end{figure}
\begin{eqnarray}
d_{\alpha }^{\text{att}} &=&X_{r}X_{\text{att}}d_{\alpha },  \notag \\
\sigma _{\alpha }^{\text{att}} &=&X_{r}(X_{\text{att}}\sigma _{\alpha }X_{%
\text{att}}^{T}+Y_{\text{att}})X_{r}^{T}+Y_{r},  \notag \\
d_{s}^{\text{att}} &=&d_{s},  \notag \\
\sigma _{s}^{\text{att}} &=&X_{r}(X_{\text{att}}\sigma _{s}X_{\text{att}%
}^{T}+Y_{\text{att}})X_{r}^{T}+Y_{r}.  \label{24}
\end{eqnarray}%
In this context, we begin by addressing a single-parameter estimation
problem involving the attenuation parameter $\vartheta .$ Based on Eqs. (\ref%
{16}) and (\ref{24}), we can derive the corresponding QFI of the attenuation
parameter for a coherent state input%
\begin{equation}
F_{\vartheta }^{\alpha }=4\left[ \frac{\left\vert \alpha \right\vert
^{2}\sin ^{2}\vartheta }{\Delta _{1}}+\frac{\Delta _{2}}{\Delta _{1}-1}%
\right] \cosh ^{2}r,  \label{25}
\end{equation}%
where $\left\vert \alpha \right\vert ^{2}=$Re($\alpha $)$^{2}+$Im($\alpha $)$%
^{2}$ is the mean photon number of the coherent state and 
\begin{eqnarray}
\Delta _{1} &=&\sinh ^{2}r+(1+2\bar{N}_{\text{att}}\sin ^{2}\vartheta )\cosh
^{2}r,  \notag \\
\Delta _{2} &=&\frac{\bar{N}_{\text{att}}^{2}\sin ^{2}(2\vartheta )}{2(1+%
\bar{N}_{\text{att}}\sin ^{2}\vartheta )}.  \label{26}
\end{eqnarray}%
In particular, in the case of $\bar{N}_{att}=0$ and $r\rightarrow 0,$ the
Eq. (\ref{25}) aligns with the previous work \cite{28}. Generally, a higher
value of QFI corresponds to greater estimation precision. To clearly see
this point, according to Eq. (\ref{25}), we plot the QFI for the input
coherent state as a function of the acceleration parameter $r,$ as shown in
Fig. 2(a). Evidently, as $r$ increases, the QFI progressively decreases and
asymptotically approaches a non-zero value in the limit of infinite
acceleration, consistent with the results presented in Ref. \cite{54}.
Further, we also consider the effects of the attenuation parameter $%
\vartheta $ on the QFI, as depicted in Fig. 2(b). It is clear that the QFI
exhibits a periodic variation with the increase of $\vartheta $. 
\begin{figure}[tbp]
\label{Fig4} \centering\includegraphics[width=0.8\columnwidth]{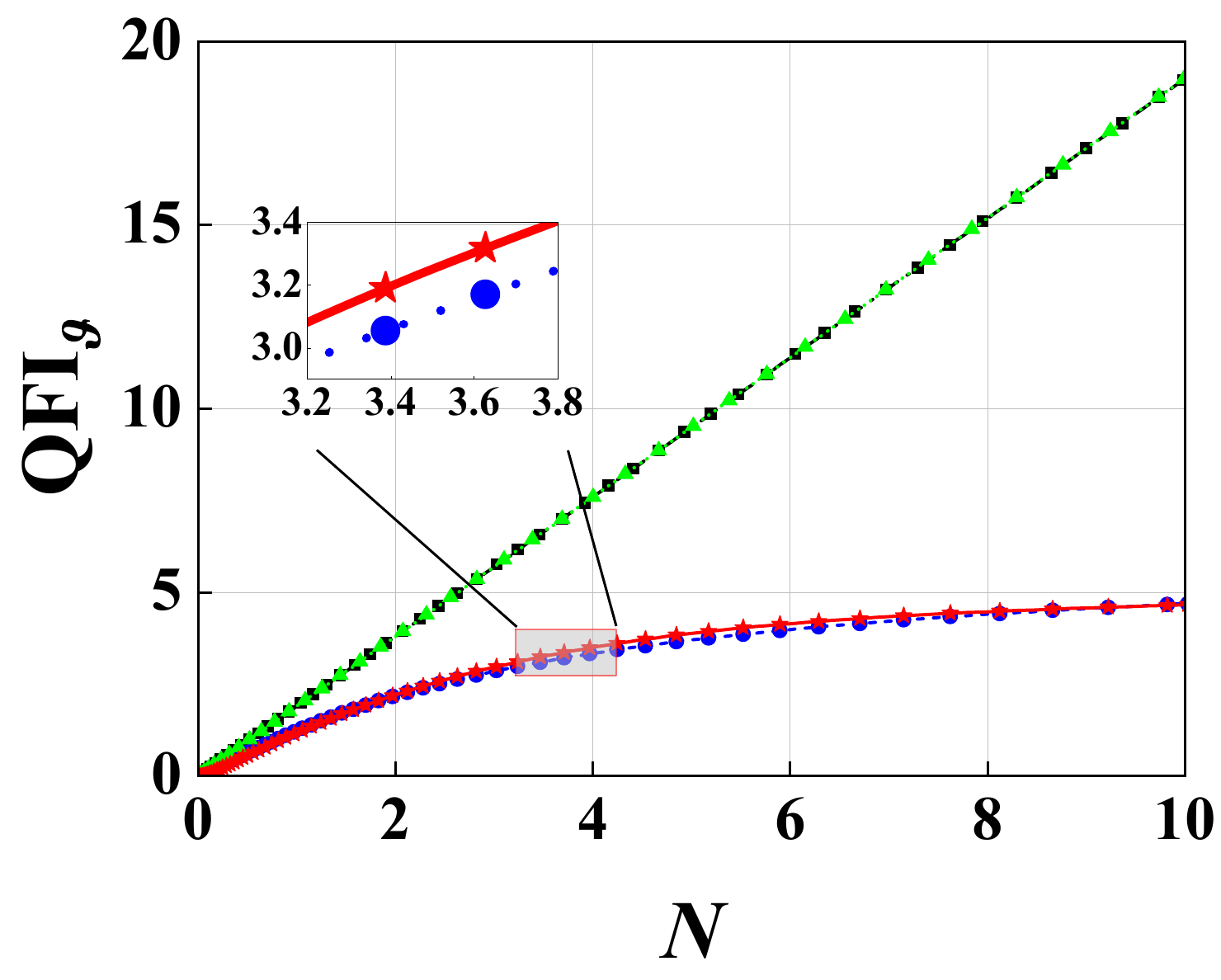}%
\newline
\caption{{}(Color online) The QFI with respect to the attenuation parameter $%
\protect\vartheta $\ is analyzed as a function of the mean photon number $N$%
\ \ with $\protect\vartheta =\protect\pi /3,$\ $r=1$\ and\ $\bar{N}_{\text{%
att}}=0.$\ The black and blue lines represent the true QFI for the input
coherent state and the squeezed vacuum state, respectively. The green and
red lines depict the numerical fitting results obtained from Eq. (\protect
\ref{27}).}
\end{figure}

Due to the complexity of deriving analytical results, for the input squeezed
vacuum state, we focus on numerical computations of the QFI of the
attenuation parameter changing with the relevant physical parameters $r$ and 
$\vartheta $, as illustrated in Fig. 3. Similar to the case of coherent
states as input states, the QFI gradually decreases with increasing $r$ and
displays a periodic dependence on $\vartheta $. A comparative analysis of
Fig. 2(a) and Fig. 3(a) reveals that the thermal mean number $\bar{N}_{\text{%
att}}$\ plays a dual role in parameter estimation. For a coherent state
input, $\bar{N}_{\text{att}}$\ typically acts as a source of decoherence
that degrades the QFI with respect to the attenuation parameter $\vartheta $%
, as evidenced by the behavior at $\vartheta =\pi /2$\ in Fig. 2(a).
However, $\bar{N}_{\text{att}}$\ can unexpectedly enhance the QFI under
specific conditions, such as the squeezed state case shown in Fig. 3(a).
This phenomenon is intrinsically linked to the attenuation parameter $%
\vartheta $. Specifically, as illustrated in Fig. 2(b), the QFI exhibits a
positive correlation with $\bar{N}_{\text{att}}$\ in specific regions, such
as within the regime $\vartheta <0.68$. This behavior highlights the complex
interplay between thermal noise and channel attenuation, suggesting that the
presence of a thermal background does not always strictly inhibit estimation
precision but can coexist with quantum resources to enhance the estimation
precision. Furthermore, we investigate the influence of various input states
on the QFI. Fig. 4 illustrates the QFI as a function of the mean photon
number $N$\ for both the coherent state and the squeezed vacuum state. Here,
the mean photon number for the coherent state and the squeezed vacuum state
is defined as $\left\vert \alpha \right\vert ^{2}$ and $\sinh ^{2}s,$
respectively. It is observed that the QFI for both cases exhibits a
monotonic increase with the mean photon number. To further quantify the
scaling behavior, we perform numerical fitting on the QFI of the attenuation
parameter, yielding the following empirical formulas%
\begin{eqnarray}
F_{\vartheta }^{\alpha } &\simeq &1.9N,  \notag \\
F_{\vartheta }^{s} &\simeq &F_{\vartheta }^{\alpha }-1.54N(1-e^{-0.18N})-0.47%
\sqrt{N},  \label{27}
\end{eqnarray}%
where $F_{\vartheta }^{\alpha }$\ and $F_{\vartheta }^{s}$\ denote the QFI
for the coherent state and the squeezed vacuum state, respectively. As shown
in Fig. 4, these empirical formulas can saturate the corresponding true QFI.%
\textbf{\ } 
\begin{figure}[tbp]
\label{Fig5} \centering\includegraphics[width=0.8\columnwidth]{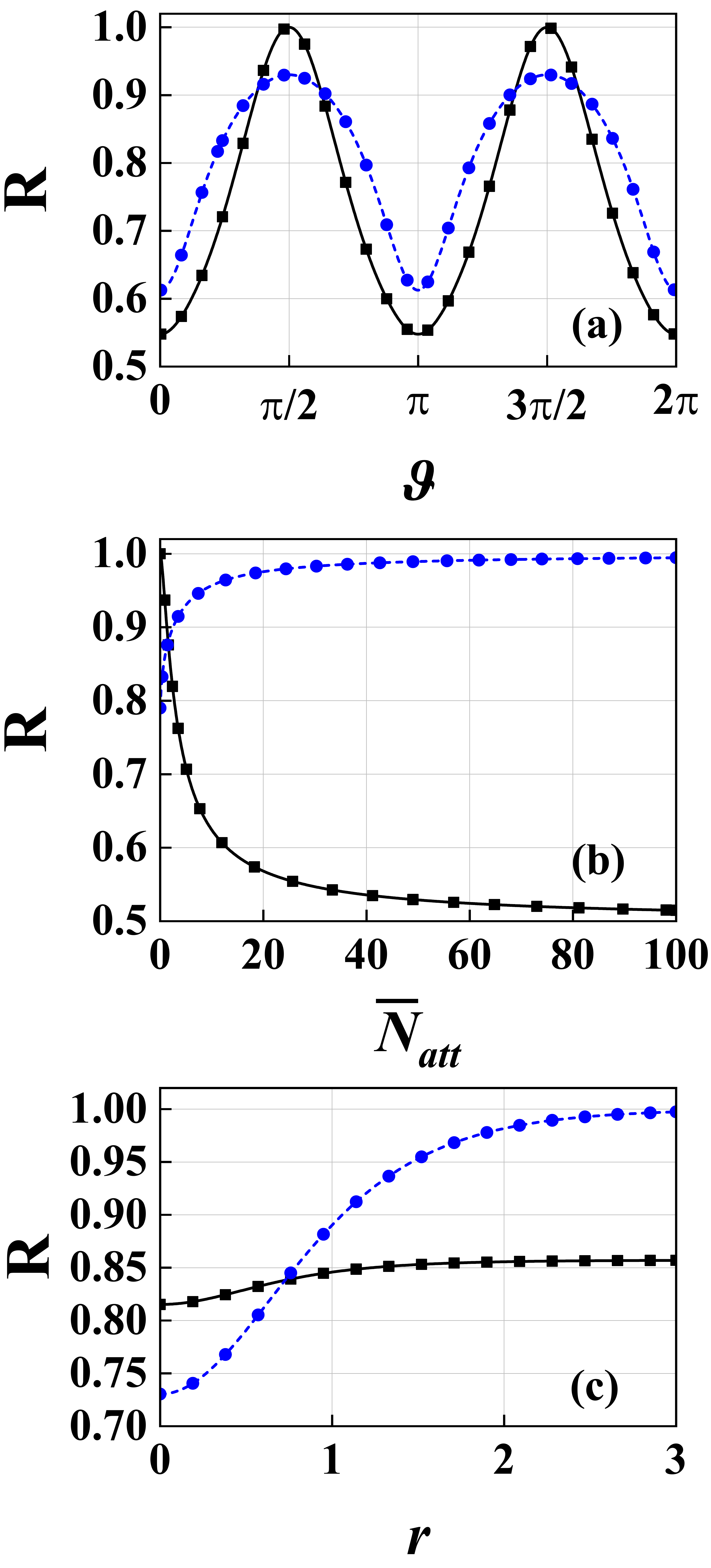}%
\newline
\caption{{}(Color online) The ratio $R$\ between the CFI and the QFI with
respect to the attenuation parameter $\protect\vartheta $\ for the input
coherent state is analyzed as a function of (a) the attenuation parameter $%
\protect\vartheta $\ with $r=1,$\ Re$(\protect\alpha )^{2}=2,$\ Im$(\protect%
\alpha )=0,$\ and $\bar{N}_{\text{att}}=2,$\ of (b) the thermal mean number $%
\bar{N}_{\text{att}}$\ with $\protect\vartheta =\protect\pi /3,$\ $r=1,$\ Re$%
(\protect\alpha )^{2}=2$, and Im$(\protect\alpha )=0,$\ and of (c) the
acceleration parameter $r$\ with $\protect\vartheta =\protect\pi /3,$\ Re$(%
\protect\alpha )^{2}=2$, Im$(\protect\alpha )=0,$\ and $\bar{N}_{\text{att}%
}=2.$\ The black and blue lines correspond to homodyne and heterodyne
measurements, respectively.}
\end{figure}
\begin{figure}[tbp]
\label{Fig6} \centering\includegraphics[width=0.8\columnwidth]{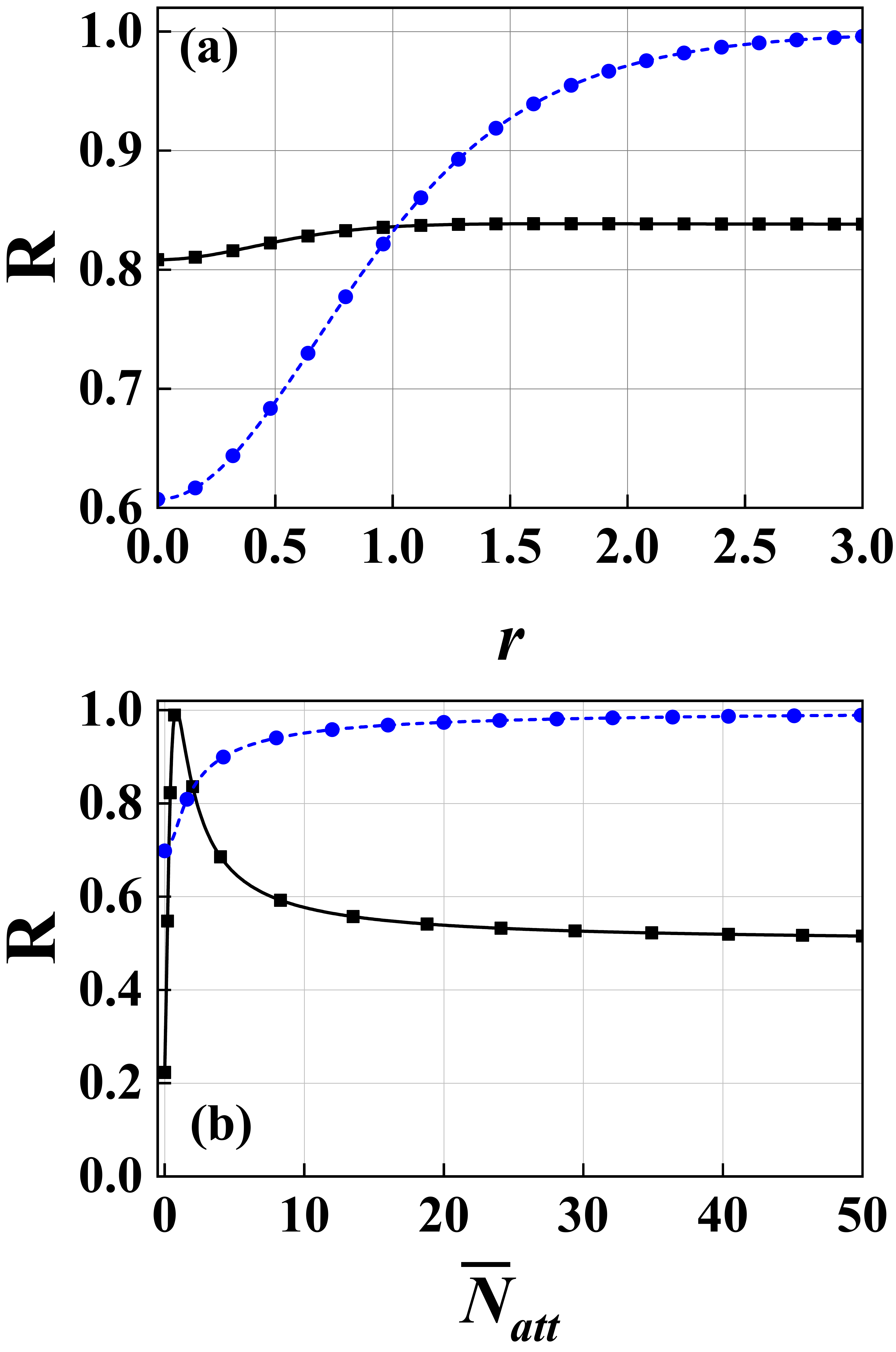}%
\newline
\caption{{}(Color online) The ratio $R$\ between the CFI and the QFI with
respect to the attenuation parameter $\protect\vartheta $\ for the input
squeezed vacuum state is analyzed as a function of (a) the acceleration
parameter $r$\ with $\protect\vartheta =\protect\pi /3,$\ $s=0.5,$\ and $%
\bar{N}_{\text{att}}=2,$\ of (b) the thermal mean number $\bar{N}_{\text{att}%
}$\ with $\protect\vartheta =\protect\pi /3,$\ $r=1,$ and\ $s=0.5.$\ The
black and blue lines correspond to homodyne and heterodyne measurements,
respectively.}
\end{figure}

Subsequently, we investigate the conditions under which the CFI for Gaussian
measurements, including homodyne and heterodyne measurements, saturates the
QFI. According to Eq. (\ref{12}), the CFI of the attenuation parameter for a
coherent state input under both measurement schemes are derived as%
\begin{eqnarray}
\tilde{F}_{\vartheta }^{\text{hom}} &=&\frac{(4\Delta _{3}\cosh ^{4}r+\text{%
Re}(\alpha )^{2}\sinh ^{2}(2r))(\sin ^{2}\vartheta )}{\Delta _{1}^{2}}, 
\notag \\
\tilde{F}_{\vartheta }^{\text{het}} &=&\frac{4\Delta _{5}\sin ^{2}\vartheta 
}{\Delta _{4}^{2}},  \label{28}
\end{eqnarray}%
\textbf{\ }where $\tilde{F}_{\vartheta }^{\text{hom}}$\ and $\tilde{F}%
_{\vartheta }^{\text{het}}$\ represent the CFI of homodyne and heterodyne
measurements, respectively, $\Delta _{1}$\ is given by Eq. (\ref{26}) and 
\textbf{\ }%
\begin{eqnarray}
\Delta _{3} &=&\bar{N}_{\text{att}}^{2}+(1+\bar{N}_{\text{att}})\text{Re}%
(\alpha )^{2}  \notag \\
&&+\bar{N}_{\text{att}}(\bar{N}_{\text{att}}-\text{Re}(\alpha )^{2})\cos
(2\vartheta ),  \notag \\
\Delta _{4} &=&2+\bar{N}_{\text{att}}-\bar{N}_{\text{att}}\cos (2\vartheta ),
\notag \\
\Delta _{5} &=&2\bar{N}_{\text{att}}^{2}+(2+\bar{N}_{\text{att}})\left\vert
\alpha \right\vert ^{2}  \notag \\
&&+\bar{N}_{\text{att}}(2\bar{N}_{\text{att}}-\left\vert \alpha \right\vert
^{2})\cos (2\vartheta ).  \label{29}
\end{eqnarray}%
Obviously, by setting $\vartheta =\pi /2$\ and Im$(\alpha )=0,$ the QFI $%
F_{\vartheta }^{\alpha }$\ given by Eq. (\ref{25}) and the CFI $\tilde{F}%
_{\vartheta }^{\text{hom}}$\ of homodyne measurement coincide as $\tilde{F}%
_{\vartheta }^{\text{hom}}=F_{\vartheta }^{\alpha }=\frac{4\text{Re}(\alpha
)^{2}\cosh ^{2}r}{\bar{N}_{\text{att}}+(1+\bar{N}_{\text{att}})\cosh (2r)}.$
Similarly, for $\bar{N}_{\text{att}}=0$\ and Im$(\alpha )=0,$ the QFI $%
F_{\vartheta }^{\alpha }$\ and the CFI $\tilde{F}_{\vartheta }^{\text{hom}}$%
\ of homodyne measurement takes the form $\tilde{F}_{\vartheta }^{\text{hom}%
}=F_{\vartheta }^{\alpha }=4$Re$(\alpha )^{2}$sech$(2r)\cosh ^{2}r\sin
^{2}\vartheta $ indicating that the homodyne measurement saturates the QFI
under these conditions. Furthermore, the CFI $\tilde{F}_{\vartheta }^{\text{%
het}}$\ for heterodyne measurement is found to asymptotically approach the
QFI in the high-acceleration limit, i.e., $\tilde{F}_{\vartheta }^{\text{het}%
}=\lim_{r\rightarrow \infty }F_{\vartheta }^{\alpha }.$\ In the large-$\bar{N%
}_{\text{att}}$\ limit $\bar{N}_{\text{att}}\rightarrow \infty ,$\ we can
get the relationship $F_{\vartheta }^{\alpha }=\tilde{F}_{\vartheta }^{\text{%
het}}=2\tilde{F}_{\vartheta }^{\text{hom}}=4\cot ^{2}\vartheta .$\ To
quantitatively assess the conditions under which a specific Gaussian
measurement reaches the ultimate precision limit, we introduce the ratio
between the CFI and the QFI as%
\begin{equation}
R=\frac{\tilde{F}_{\vartheta }}{F_{\vartheta }}.  \label{30}
\end{equation}%
From Eq. (\ref{30}), the value of $R$ is always less than or equal to 1. A
value of $R=1$\ signifies that the chosen Gaussian measurement can saturate
the QFI. To clearly see this point, we plot the ratio $R$\ as a function of
the attenuation parameter $\vartheta ,$ as shown in Fig. 5(a). For homodyne
measurement, the ratio $R$\ reaches its theoretical maximum $1$\ precisely
at $\vartheta =\pi /2$\ and $\vartheta =3\pi /2,$ indicating that homodyne
measurement is an optimal measurement scheme that fully saturates the QFI
under these specific conditions. However, its efficiency exhibits
significant variations and drops to its minimum as the attenuation parameter
deviates from these optimal points. In contrast, heterodyne measurement
maintains a consistently sub-optimal performance across the entire range of $%
\vartheta $, with the ratio $R$\ always remaining below $1$. Notably, while
it fails to saturate the QFI, the heterodyne measurement displays a smoother
and more robust response with smaller fluctuations, offering superior
precision in the regions where the efficiency of homodyne measurement is at
its lowest. In Fig. 5(b), we consider the impact of the thermal mean number $%
\bar{N}_{\text{att}}$ on the ratio $R$. It is observed that in the absence
of thermal noise ($\bar{N}_{\text{att}}=0$), homodyne measurement is the
superior scheme, successfully saturating the QFI with $R=1$. However, as the
thermal noise increases, the efficiency of homodyne measurement undergoes a
sharp decline, asymptotically approaching a value of $0.5$\ in the
high-attenuation limit. Conversely, the performance of heterodyne
measurement improves significantly with rising thermal noise; while it is
sub-optimal in the low-noise regime, its ratio $R$\ monotonically increases
and eventually approaches $1$\ as $\bar{N}_{\text{att}}$\ becomes large.
This behavior demonstrates a clear crossover between the two schemes,
highlighting that heterodyne measurement becomes the more efficient and even
optimal measurement choice in the presence of strong thermal background
noise, whereas homodyne measurement is better suited for low-noise
environments. In Fig. 5(c), we examine the dependence of the ratio $R$\ on
the acceleration parameter $r$\ to evaluate how non-inertial effects
influence the estimation precision. For heterodyne measurement, the ratio $R$%
\ exhibits a monotonic increase as the acceleration parameter $r$\ grows,
asymptotically approaching $1$\ in the high-acceleration regime. This trend
indicates that heterodyne measurement\textbf{\ }becomes an optimal
measurement scheme capable of fully saturating the QFI. Conversely, while
the efficiency of homodyne measurement $R$\ also shows an initial upward
trend with $r$, it quickly reaches a plateau at a sub-optimal level,
remaining significantly below the QFI. 
\begin{figure}[tbp]
\label{Fig7} \centering\includegraphics[width=0.8\columnwidth]{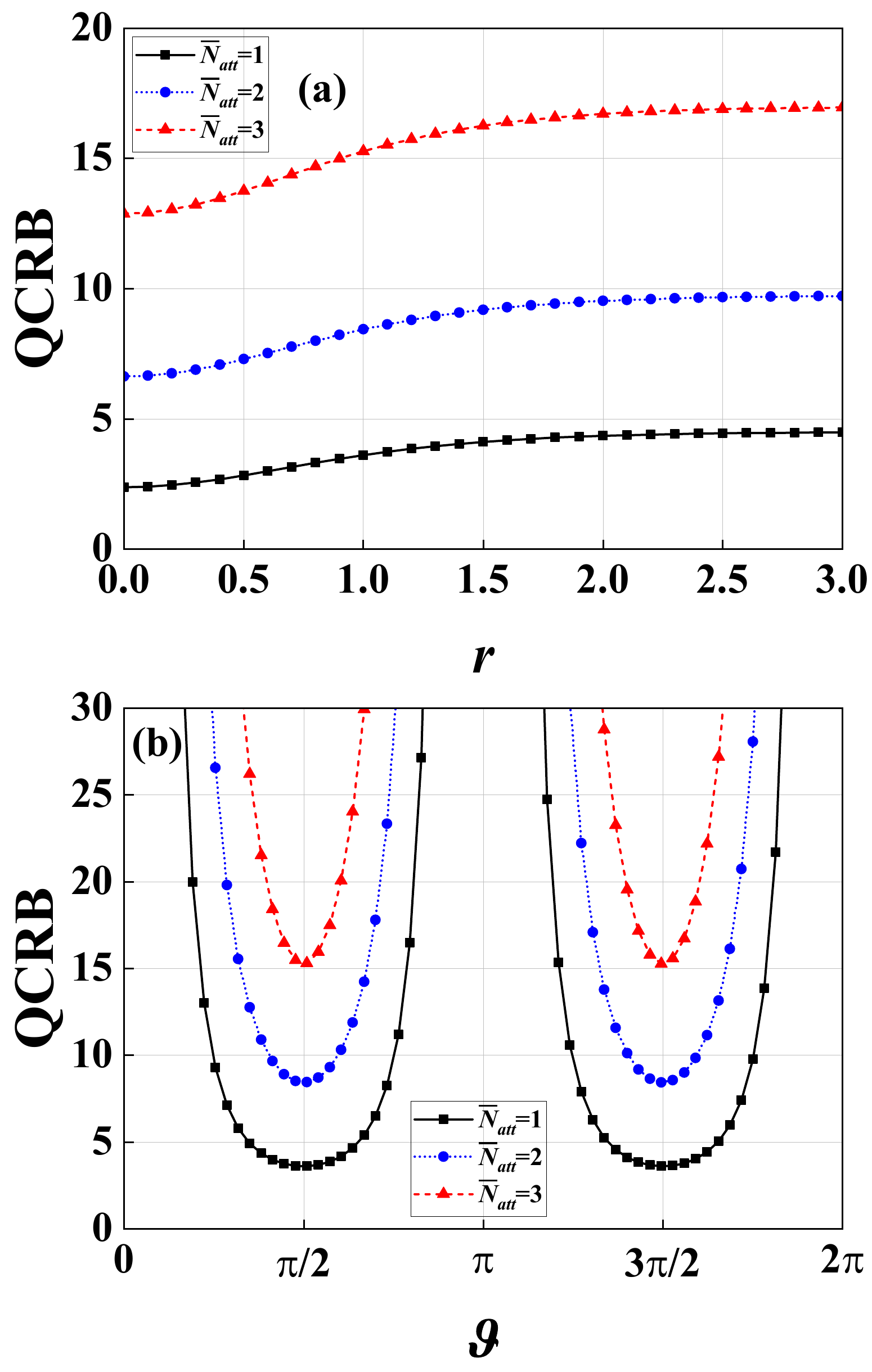}%
\newline
\caption{{}(Color online) The QCRB with respect to the two-parameter
estimation including the attenuation parameter $\protect\vartheta $\ and the
thermal mean number $\bar{N}_{\text{att}}$\ of the thermal attenuator
channel for the input coherent state is analyzed as a function of (a) the
acceleration parameter $r$\ with $\protect\vartheta =\protect\pi /2$ and\ $%
\left\vert \protect\alpha \right\vert ^{2}=2,$ and of (b) the attenuation
parameter $\protect\vartheta $\ with $r=1$ and\ $\left\vert \protect\alpha %
\right\vert ^{2}=2.$}
\end{figure}
\begin{figure}[tbp]
\label{Fig8} \centering\includegraphics[width=0.8\columnwidth]{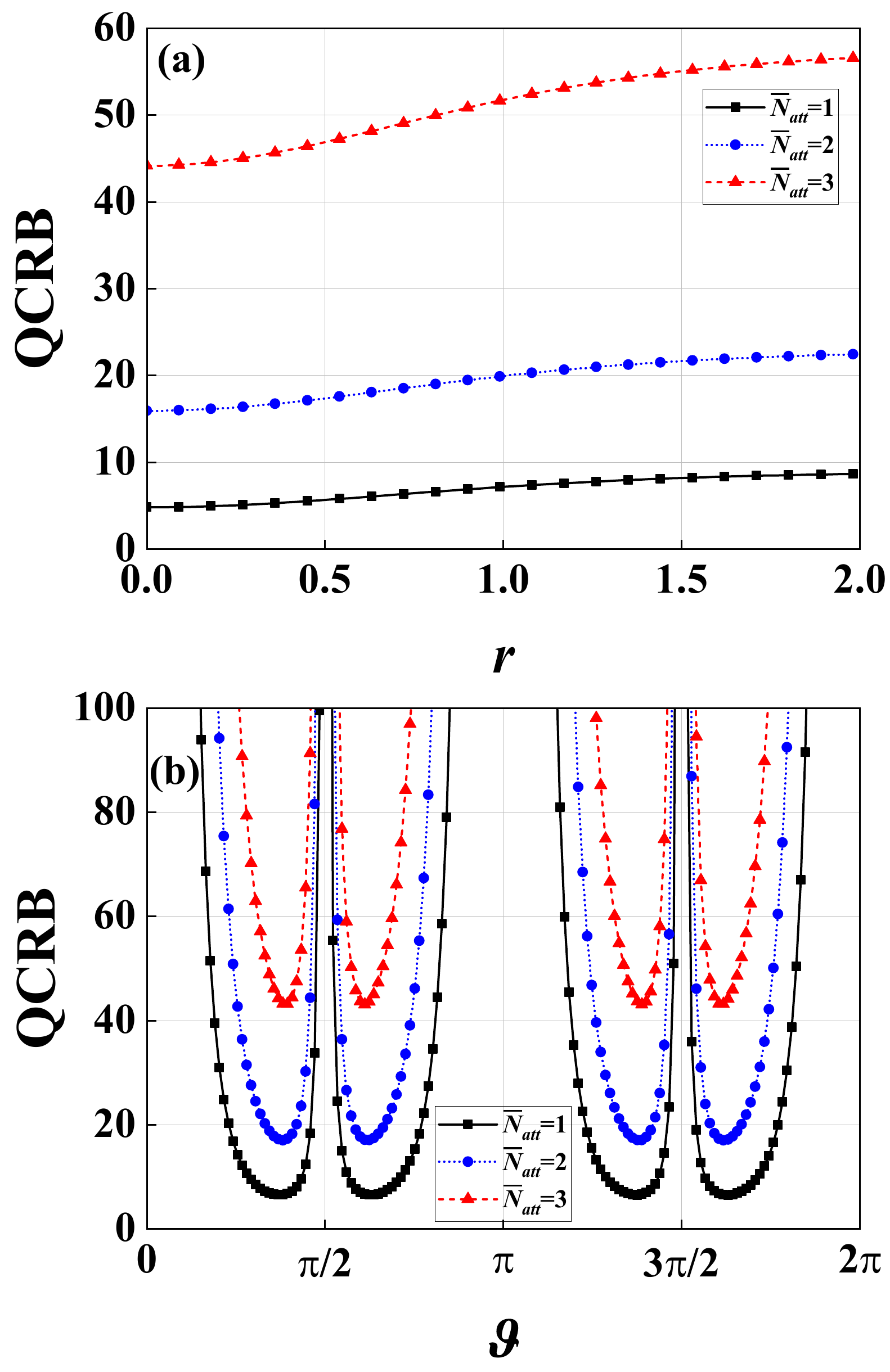}%
\newline
\caption{{}(Color online) The QCRB with respect to the two-parameter
estimation including the attenuation parameter $\protect\vartheta $\ and the
thermal mean number $\bar{N}_{\text{att}}$\ of the thermal attenuator
channel for the input squeezed vacuum state is analyzed as a function of (a)
the acceleration parameter $r$\ with $\protect\vartheta =\protect\pi /3$
and\ $s=1,$ and of (b) the attenuation parameter $\protect\vartheta $\ with $%
r=1$ and\ $s=1.$}
\end{figure}

For the input squeezed vacuum state, we mainly focus on numerical
computations of the ratio $R$\ between the CFI and the QFI changing with the
relevant physical parameters $r$\ and $\bar{N}_{\text{att}}$, as shown in
Fig. 6. In Fig. 6(a), the ratio $R$ for heterodyne measurement\ demonstrates
a significant monotonic increase with the acceleration parameter $r$,
starting from a relatively low value in the small acceleration regime and
asymptotically approaching $1$\ as $r$\ increases. This indicates that
heterodyne measurement becomes an optimal measurement scheme that fully
saturates the QFI in the high-acceleration limit. In contrast, the ratio $R$
for homodyne measurement shows a much more gradual enhancement and quickly
enters a plateau, remaining at a sub-optimal level significantly below the
theoretical maximum. The crossover between the two curves further suggests
that while homodyne measurement may offer a higher information ratio at very
low accelerations, heterodyne measurement becomes far more effective as the
non-inertial effects intensify, eventually capturing nearly all the
available quantum information at large $r$. In Fig. 6(b), the ratio $R$ for
homodyne measurement\ exhibits a striking non-monotonic behavior in the
low-noise regime; it initially increases sharply to reach a peak of 1, where
the QFI is fully saturated, before undergoing a steady decline toward an
asymptotic value of $0.5$\ as the thermal noise intensifies. In contrast,
the ratio $R$\ for heterodyne measurement shows a consistent monotonic
improvement with increasing $\bar{N}_{\text{att}}$. Although heterodyne
measurement is sub-optimal when the thermal background is weak, its ratio
asymptotically approaches 1 in the limit of large thermal mean number,
eventually outperforming homodyne measurement. This crossover highlights
that while homodyne measurement can be exceptionally precise under specific
low-noise conditions, heterodyne detection provides a more reliable and
ultimately optimal approach for parameter estimation in strongly dissipative
thermal environments. 
\begin{figure}[tbp]
\label{Fig9} \centering\includegraphics[width=0.8\columnwidth]{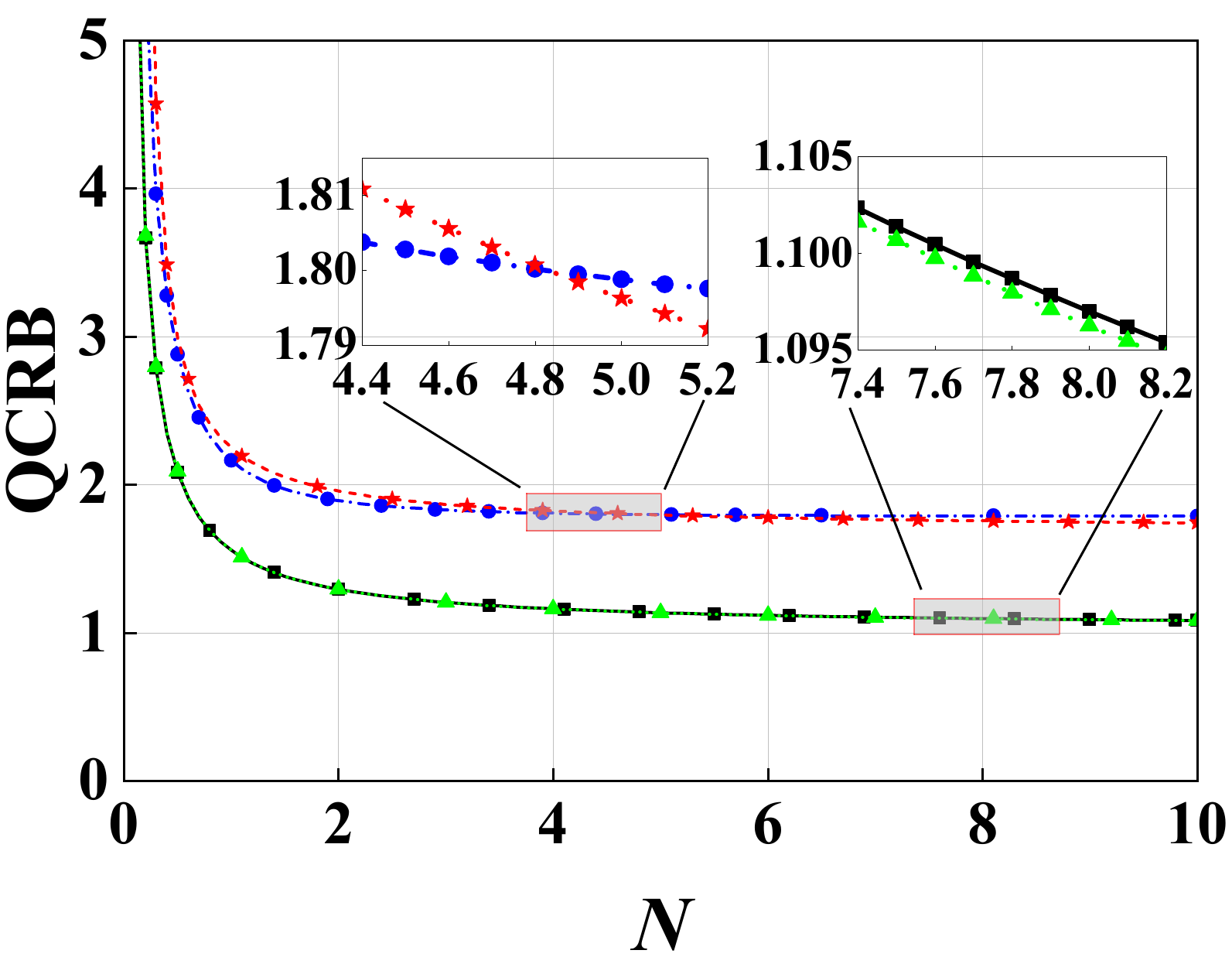}%
\newline
\caption{{}(Color online) The QCRB with respect to the two-parameter
estimation including the attenuation parameter $\protect\vartheta $\ and the
thermal mean number $\bar{N}_{\text{att}}$\ of the thermal attenuator
channel\ is analyzed as a function of the mean photon number $N$\ \ with $%
\protect\vartheta =\protect\pi /3,$\ $r=1$\ and\ $\bar{N}_{\text{att}}=0.$\
The black and blue lines represent the true QCRB for the input coherent
state and the squeezed vacuum state, respectively. The green and red lines
denote the numerical fitting results given by Eq. (\protect\ref{34}).}
\end{figure}
\textbf{\ }

Then, we analyze a two-parameter estimation problem including the
attenuation parameter $\vartheta $ and the thermal mean number $\bar{N}_{%
\text{att}}$ of the thermal attenuator channel. For a coherent state input,
according to Eqs. (\ref{19}) and (\ref{24}), the corresponding commutation
relation between symmetric logarithmic derivative operators $\hat{L}%
_{\vartheta }$ and $\hat{L}_{\bar{N}_{\text{att}}}$ can be expressed as%
\begin{equation}
\lbrack \hat{L}_{\vartheta },\hat{L}_{\bar{N}_{\text{att}}}]=2i\left[
(\Gamma _{1}+\Gamma _{3})\hat{x}_{1}+(\Gamma _{2}+\Gamma _{4})\hat{p}_{1}%
\right] ,  \label{31}
\end{equation}%
where $\Gamma _{1},$ $\Gamma _{2},$ $\Gamma _{3},$ and $\Gamma _{4}$ are
shown in Appendix A. Obviously, the symmetric logarithmic derivative
operators $\hat{L}_{\vartheta }$ and $\hat{L}_{\bar{N}_{\text{att}}}$ are
non-commutative. This further suggests that it is impossible to establish a
shared set of eigenstates for $\hat{L}_{\vartheta }$ and $\hat{L}_{\bar{N}_{%
\text{att}}}$, which constitute the optimal measurement scheme necessary for
attaining the QCRB through single-copy measurements. However, it is notable
that the corresponding mean Uhlmann curvature matrix is observed to be a
zero matrix. This implies that the QCRB serves as an asymptotically tight
precision limit. In this scenario, we obatin the QCRB with respect to the
two-parameter estimation including the attenuation parameter $\vartheta $
and the thermal mean number $\bar{N}_{\text{att}}$ of the thermal attenuator
channel for the input coherent state 
\begin{equation}
Q_{\text{att}}^{\alpha }=\frac{[\Upsilon _{1}+(\Upsilon _{2}+\Upsilon
_{3})\Upsilon _{4}]\Upsilon _{5}}{\Upsilon _{6}},  \label{32}
\end{equation}%
where%
\begin{eqnarray}
\Upsilon _{1} &=&8\left\vert \alpha \right\vert ^{2}[\bar{N}_{\text{att}%
}\cos (2\vartheta )-\bar{N}_{\text{att}}-2],  \notag \\
\Upsilon _{2} &=&\cos (2\vartheta )[1+2\bar{N}_{\text{att}}(\left\vert
\alpha \right\vert ^{2}-2\bar{N}_{\text{att}})],  \notag \\
\Upsilon _{3} &=&-2\left\vert \alpha \right\vert ^{2}(2+\bar{N}_{\text{att}%
})-4\bar{N}_{\text{att}}^{2}-1,  \notag \\
\Upsilon _{4} &=&2[2\bar{N}_{\text{att}}\cos (2\vartheta )\cosh ^{2}r  \notag
\\
&&-(2+\bar{N}_{\text{att}})\cosh (2r)-\bar{N}_{\text{att}}],  \notag \\
\Upsilon _{5} &=&\csc ^{4}\theta \lbrack \bar{N}_{\text{att}}\cos
(2\vartheta )+\text{sech}^{2}r  \notag \\
&&-\tanh ^{2}r-\bar{N}_{\text{att}}-1],  \notag \\
\Upsilon _{6} &=&16\left\vert \alpha \right\vert ^{2}[2\bar{N}_{\text{att}%
}\cos (2\vartheta )\cosh ^{2}r  \notag \\
&&-(2+\bar{N}_{\text{att}})\cosh (2r)+2-\bar{N}_{\text{att}}].  \label{33}
\end{eqnarray}

In order to visually see the impacts of the Unruh effect on the QCRB with
respect to the two-parameter estimation including the attenuation parameter $%
\vartheta $ and the thermal mean number $\bar{N}_{\text{att}}$ of the
thermal attenuator channel, we show the QCRB with respect to the
acceleration parameter $r$ in Fig. 7(a). It is evident that as $r$
increases, the QCRB gradually rises and asymptotically converges to a
non-zero value in the limit of infinite acceleration. Furthermore, the QCRB
for the case of $\bar{N}_{\text{att}}=1$ exhibits better estimation
performance comepared with $\bar{N}_{\text{att}}=2$ and $\bar{N}_{\text{att}%
}=3.$ Additionally, we also examine the influence of the attenuation
parameter $\vartheta $ on the QCRB, as illustrated in Fig. 7(b). Our
analysis reveals that the QCRB undergoes periodic fluctuations with
increasing $\vartheta ,$ achieving optimal estimation performance at $%
\vartheta =\pi /2$ and $\vartheta =3\pi /2$.

Likewise, for the input squeezed vacuum state, it is also evident that the
symmetric logarithmic derivative operators $\hat{L}_{\vartheta }$ and $\hat{L%
}_{\bar{N}_{\text{att}}}$ are non-commutative. Nevertheless, the
corresponding mean Uhlmann curvature matrix is found to be a zero matrix,
indicating that the QCRB with respect to the two-parameter estimation
including the attenuation parameter $\vartheta $ and the thermal mean number 
$\bar{N}_{\text{att}}$ of the thermal attenuator channel is an
asymptotically tight precision limit. Owing to the intricate nature of
deriving analytical results, we concentrate on numerical computations of the
QCRB as functions of the relevant physical parameters $r$ and $\vartheta $,
as shown in Fig. 8. Consistent with the scenario where coherent states serve
as input states, the QCRB progressively increases with rising $r$ and
exhibits a periodic dependence on $\vartheta $. Moreover, we examine the
influence of different input states on the QCRB. Fig. 9 illustrates the QCRB
as a function of the mean photon number $N$\ for both the coherent state and
the squeezed vacuum state. It is observed that for both cases, the QCRB
exhibits a monotonic decrease as $N$\ increases, indicating a consistent
enhancement in estimation precision with more input resources. To further
characterize the scaling behavior of these precision limits, we perform a
numerical fit on the QCRB with respect to the two-parameter estimation
including the attenuation parameter $\vartheta $\ and the thermal mean
number $\bar{N}_{\text{att}}$\ of the thermal attenuator channel, yielding
the following empirical formulas%
\begin{eqnarray}
Q_{\text{att}}^{\alpha } &\simeq &1.03+\frac{0.53}{N},  \notag \\
Q_{\text{att}}^{s} &\simeq &Q_{\text{att}}^{\alpha }+\frac{0.03}{N^{3}}+0.66,
\label{34}
\end{eqnarray}%
where $Q_{\text{att}}^{\alpha }$\ and $Q_{\text{att}}^{s}$\ correspond to
the QCRB for the coherent state and the squeezed vacuum state, respectively.
As shown in Fig. 9, these empirical formulas can saturate the corresponding
true QCRB. 
\begin{figure}[tbp]
\label{Fig10} \centering\includegraphics[width=0.8\columnwidth]{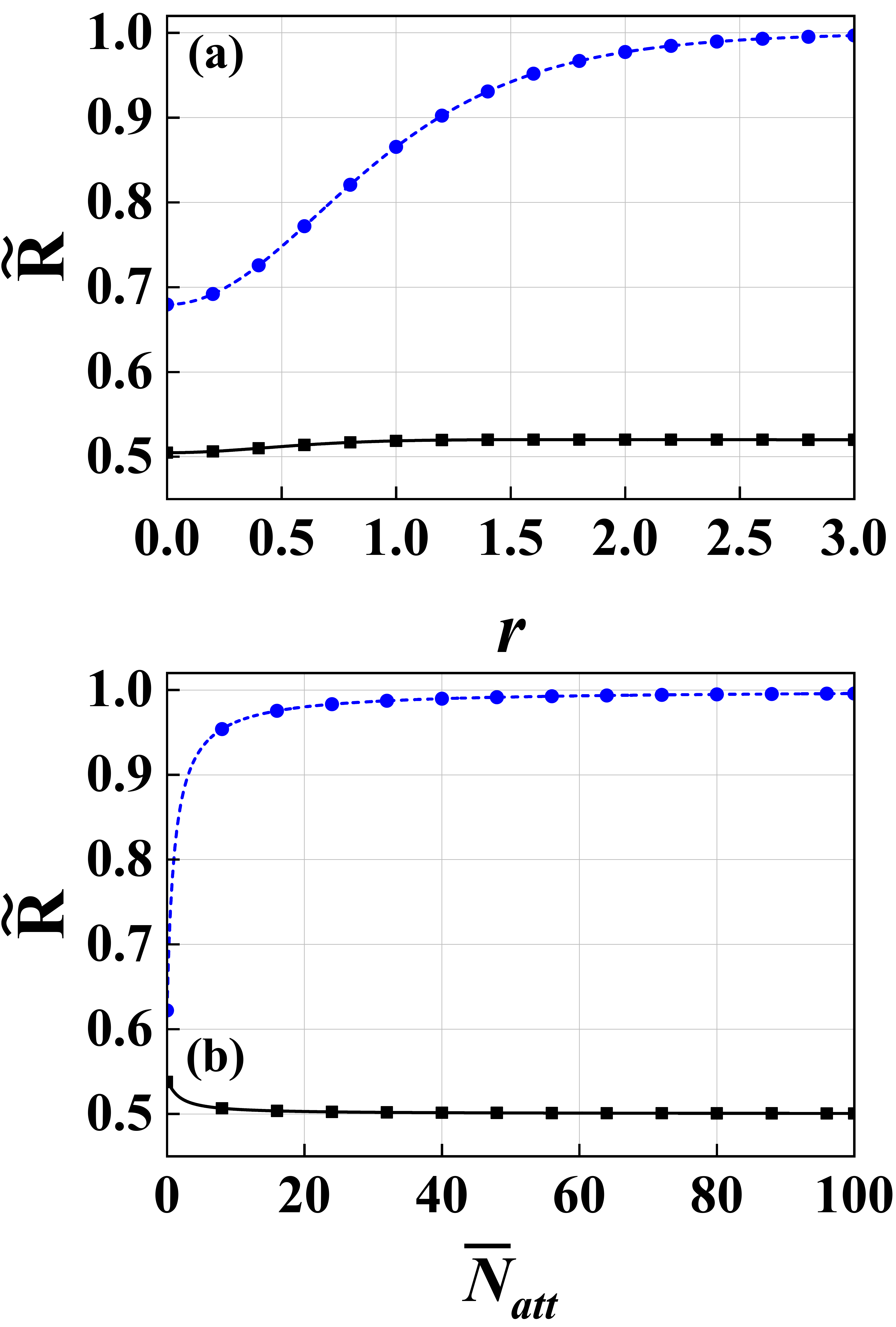}%
\newline
\caption{{}(Color online) The ratio $\tilde{R}$\ between the QCRB and the
CRB with respect to the two-parameter estimation including the attenuation
parameter $\protect\vartheta $\ and the thermal mean number $\bar{N}_{\text{%
att}}$\ of the thermal attenuator channel for the input coherent state is
analyzed as a function of (a) the acceleration parameter $r$\ with $\protect%
\vartheta =\protect\pi /2,$\ Re$(\protect\alpha )^{2}=2,$\ Im$(\protect%
\alpha )=0,$\ and $\bar{N}_{\text{att}}=2,$\ of (b) the thermal mean number $%
\bar{N}_{\text{att}}$\ with $\protect\vartheta =\protect\pi /2,$\ $r=1,$\ Re$%
(\protect\alpha )^{2}=2,$\ and Im$(\protect\alpha )=0.$ The black and blue
lines correspond to homodyne and heterodyne measurements, respectively. }
\end{figure}
\begin{figure}[tbp]
\label{Fig11} \centering\includegraphics[width=0.8\columnwidth]{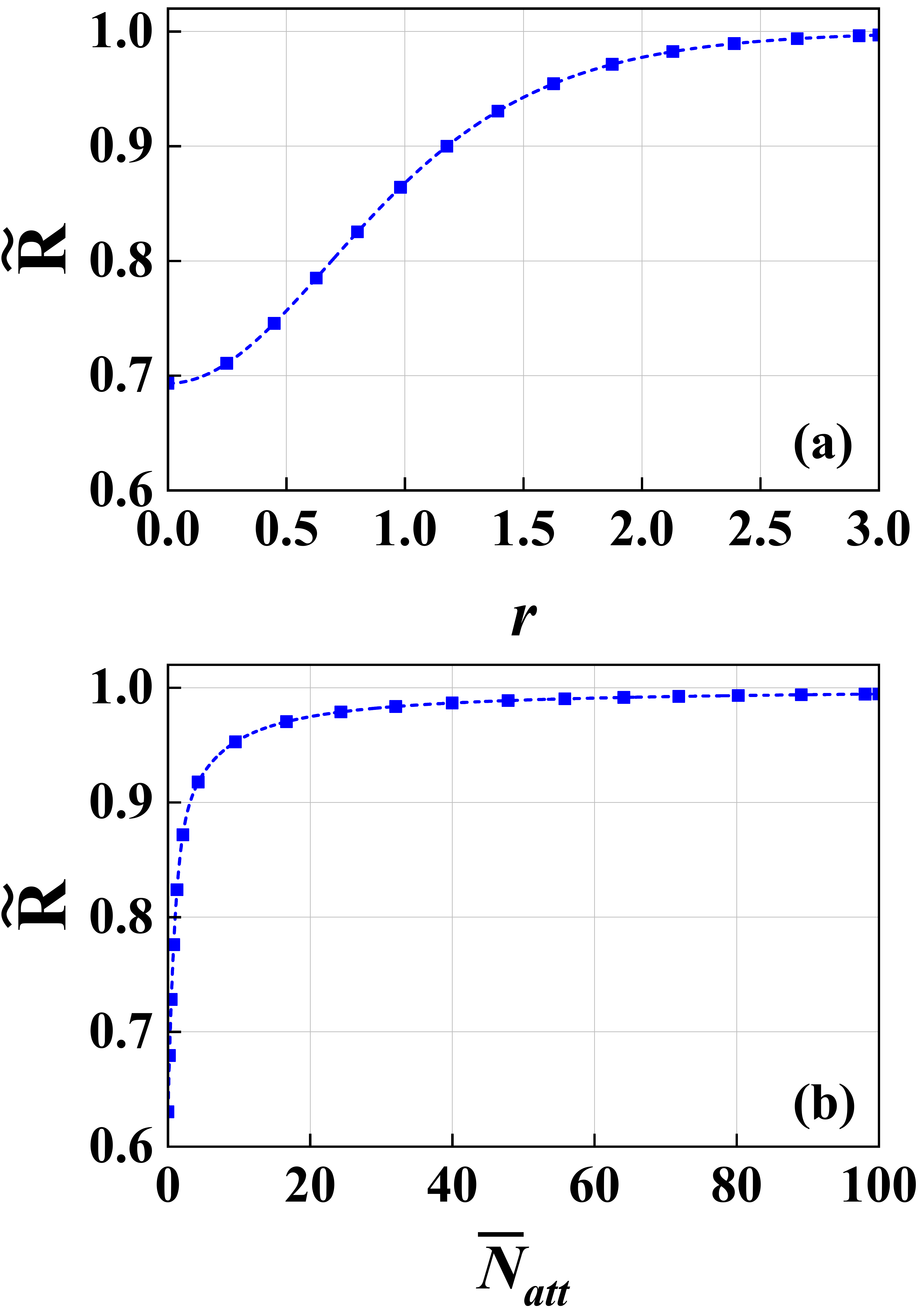}%
\newline
\caption{{}(Color online) The ratio $\tilde{R}$\ between the QCRB and the
CRB with respect to the two-parameter estimation including the attenuation
parameter $\protect\vartheta $\ and the thermal mean number $\bar{N}_{\text{%
att}}$\ of the thermal attenuator channel for the input squeezed vacuum
state is analyzed as a function of (a) the acceleration parameter $r$\ with $%
\protect\vartheta =\protect\pi /3,$\ $s=1,$ and $\bar{N}_{\text{att}}=2,$ of
(b) the thermal mean number $\bar{N}_{\text{att}}$\ with $\protect\vartheta =%
\protect\pi /3,$\ $r=1,$ and $s=1.$ The homodyne measurement fails to
simultaneously estimate $\protect\vartheta $\ and $\bar{N}_{\text{att}}$\
due to the singularity of the classical Fisher information matrix, which
results in a divergent CRB. Consequently, we exclude the homodyne
measurement results and focus our discussion on the performance of the
heterodyne measurement. }
\end{figure}

We now explore the potential for Gaussian measurements, such as homodyne and
heterodyne measurements, to attain the ultimate precision limit defined by
the QCRB. Utilizing the framework established in Eq. (\ref{12}), the
corresponding CRB with respect to the two-parameter estimation including the
attenuation parameter $\vartheta $\ and the thermal mean number $\bar{N}_{%
\text{att}}$\ of the thermal attenuator channel for a coherent state input
are determined to be%
\begin{eqnarray}
C_{\text{att}}^{\text{hom}} &=&(\Upsilon _{7}\Upsilon _{8}+\Upsilon
_{9})\Upsilon _{10},  \notag \\
C_{\text{att}}^{\text{het}} &=&\Upsilon _{11}\Upsilon _{12},  \label{35}
\end{eqnarray}%
where $C_{\text{att}}^{\text{hom}}$\ and $C_{\text{att}}^{\text{het}}$\ are
the CRB for homodyne and heterodyne measurements, respectively, and%
\begin{eqnarray}
\Upsilon _{7} &=&\bar{N}_{\text{att}}^{2}+\cosh (2r)(\bar{N}_{\text{att}%
}^{2}+\text{Re}(\alpha )^{2}),  \notag \\
\Upsilon _{8} &=&2\text{sech}^{2}r\csc ^{2}\vartheta ,  \notag \\
\Upsilon _{9} &=&1-4\bar{N}_{\text{att}}(\bar{N}_{\text{att}}-\text{Re}%
(\alpha )^{2}),  \notag \\
\Upsilon _{10} &=&\frac{(1+2\bar{N}_{\text{att}}\sin ^{2}\vartheta +\tanh
^{2}r)\csc ^{2}\vartheta }{4\text{Re}(\alpha )^{2}},  \notag \\
\Upsilon _{11} &=&1+2\bar{N}_{\text{att}}(\left\vert \alpha \right\vert
^{2}-2\bar{N}_{\text{att}})  \notag \\
&&+2(\left\vert \alpha \right\vert ^{2}+2\bar{N}_{\text{att}})\csc
^{2}\vartheta ,  \notag \\
\Upsilon _{12} &=&\frac{\bar{N}_{\text{att}}+\csc ^{2}\vartheta }{%
2\left\vert \alpha \right\vert ^{2}}.  \label{36}
\end{eqnarray}%
In particular, the CRB $C_{\text{att}}^{\text{het}}$\ for heterodyne
measurement is found to approach the QCRB in the high-acceleration limit,
i.e., $C_{\text{att}}^{\text{het}}=\lim_{r\rightarrow \infty }Q_{\text{att}%
}^{\alpha }.$ To evaluate the conditions under which a Gaussian measurement
reaches the ultimate precision limit, we define the ratio of the QCRB to the
CRB as%
\begin{equation}
\tilde{R}=\frac{Q_{\text{att}}^{\alpha }}{C_{\text{att}}},  \label{37}
\end{equation}%
By definition, the ratio $\tilde{R}$\ is bounded such that $\tilde{R}\leq 1$%
. A value of $\tilde{R}=1$\ means that the chosen Gaussian measurement can
saturate the QCRB, reaching the ultimate theoretical limit of precision. To
illustrate this transition and the conditions for optimality, we plot the
ratio $\tilde{R}$\ as a function of the acceleration parameter $r,$ as shown
in Fig. 10(a). For heterodyne measurement, the ratio $\tilde{R}$\ exhibits a
significant monotonic increase as $r$\ grows, starting from approximately $%
0.7$\ and asymptotically approaching the theoretical maximum $1$\ in the
high-acceleration limit. This behavior confirms that heterodyne measurement
becomes an optimal scheme capable of fully saturating the QCRB as
non-inertial effects intensify. In contrast, the efficiency of homodyne
measurement remains at a markedly sub-optimal level, plateauing near $0.5$\
with negligible improvement across the entire range of $r$. The widening gap
between the two curves illustrates that while homodyne measurement is
fundamentally limited in this regime, heterodyne measurement is uniquely
effective at extracting the available quantum information as the
acceleration increases. In Fig. 10(b), we analyze the ratio $\tilde{R}$\ as
a function of the thermal mean number $\bar{N}_{\text{att}}$. For heterodyne
detection, the ratio $\tilde{R}$\ exhibits a rapid increase as the value of $%
\bar{N}_{\text{att}}$\ grows, asymptotically reaching $1$. This indicates
that heterodyne measurement is an optimal measurement scheme, successfully
saturating the QCRB in the high-$\bar{N}_{\text{att}}$\ regime. In contrast,
the efficiency of homodyne measurement $\tilde{R}$\ decreases sharply and
levels off at approximately $0.5$.

For the input squeezed vacuum state, we perform numerical analyses to
evaluate how the ratio $\tilde{R}$\ evolves with respect to the acceleration
parameter $r$\ and the thermal mean number $\bar{N}_{\text{att}}$, as
illustrated in Fig. 11. Our analysis reveals that the homodyne measurement
fails to simultaneously estimate $\vartheta $\ and $\bar{N}_{\text{att}}$\
due to the singularity of the classical Fisher information matrix, resulting
in a divergent CRB. Consequently, we omit the homodyne measurement results
and restrict our discussion in Fig. 11 to the performance of heterodyne
measurement. In Fig. 11(a), the ratio $\tilde{R}$\ exhibits a significant
and monotonic increase with $r$, starting from a sub-optimal value around $%
0.7$\ at the inertial limit $(r\rightarrow 0)$. As the acceleration effects
become pronounced, $\tilde{R}$\ rapidly climbs and asymptotically approaches
the theoretical maximum $1$. Fig. 11(b) illustrates the numerical results
for the ratio $\tilde{R}$\ as a function of the thermal mean number $\bar{N}%
_{\text{att}}$. The ratio $\tilde{R}$\ displays a rapid and steep increase
in the low-$\bar{N}_{\text{att}}$\ regime, starting from approximately $0.63$%
\ and swiftly crossing the $0.9$\ threshold as $\bar{N}_{\text{att}}$\
exceeds $10$. As $\bar{N}_{\text{att}}$\ continues to increase, the curve
levels off and asymptotically approaches the theoretical optimum $1$. This
trend demonstrates that for an input squeezed vacuum state, heterodyne
measurement is highly efficient and asymptotically reaches the QCRB as the
thermal mean number increases.

\subsection{Thermal amplifier channel}

Next, let us consider the influence of the Unruh effect on the estimation
precision of thermal amplifier channel parameters. Likewise, we employ the
same initial input states, i.e., the coherent state and the squeezed vacuum
state. These two quantum states are then processed through a thermal
amplifier channel, resulting in two distinct probe states, which are
influenced by the amplifier parameter $\epsilon $ and the thermal mean
number $\bar{N}_{\text{amp}}$ of the thermal amplifier channel$.$ Positing
Alice can prepare these probe states within an inertial frame, the quantum
states observed by Rob are primarily shaped by the bosonic amplification
channel, a direct consequence of the Unruh effect, as detailed in Eq. (\ref%
{21}). The entire evolution of the system is described by%
\begin{eqnarray}
d_{\alpha }^{\text{amp}} &=&X_{r}X_{\text{amp}}d_{\alpha },  \notag \\
\sigma _{\alpha }^{\text{amp}} &=&X_{r}(X_{\text{amp}}\sigma _{\alpha }X_{%
\text{amp}}^{T}+Y_{\text{amp}})X_{r}^{T}+Y_{r},  \notag \\
d_{s}^{\text{amp}} &=&d_{s},  \notag \\
\sigma _{s}^{\text{amp}} &=&X_{r}(X_{\text{amp}}\sigma _{s}X_{\text{amp}%
}^{T}+Y_{\text{amp}})X_{r}^{T}+Y_{r}.  \label{38}
\end{eqnarray}%
Based on Eqs. (\ref{19}) and (\ref{38}), we can derive the QFI of the
amplifier parameter $\epsilon $ for the input coherent state 
\begin{figure}[tbp]
\label{Fig12} \centering\includegraphics[width=0.8\columnwidth]{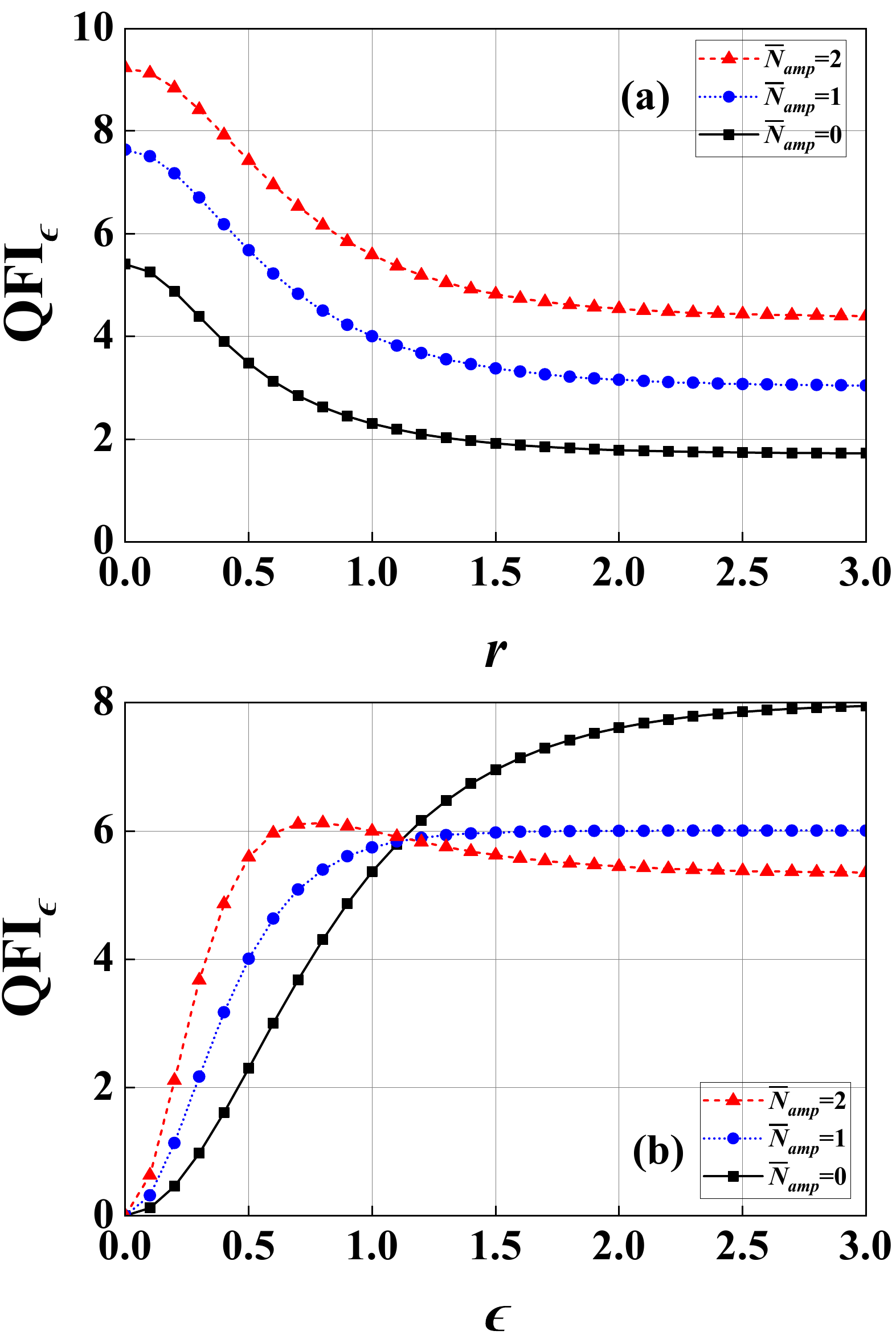}%
\newline
\caption{{}(Color online) The QFI with respect to the amplifier parameter $%
\protect\epsilon $ for the input coherent state is analyzed as a function of
(a) the acceleration parameter $r$\ with $\protect\epsilon =0.5$ and\ $%
\left\vert \protect\alpha \right\vert ^{2}=2,$ and of (b) the amplifier
parameter $\protect\epsilon $\ with $r=1$ and\ $\left\vert \protect\alpha %
\right\vert ^{2}=2.$}
\end{figure}
\begin{figure}[tbp]
\label{Fig13} \centering\includegraphics[width=0.8\columnwidth]{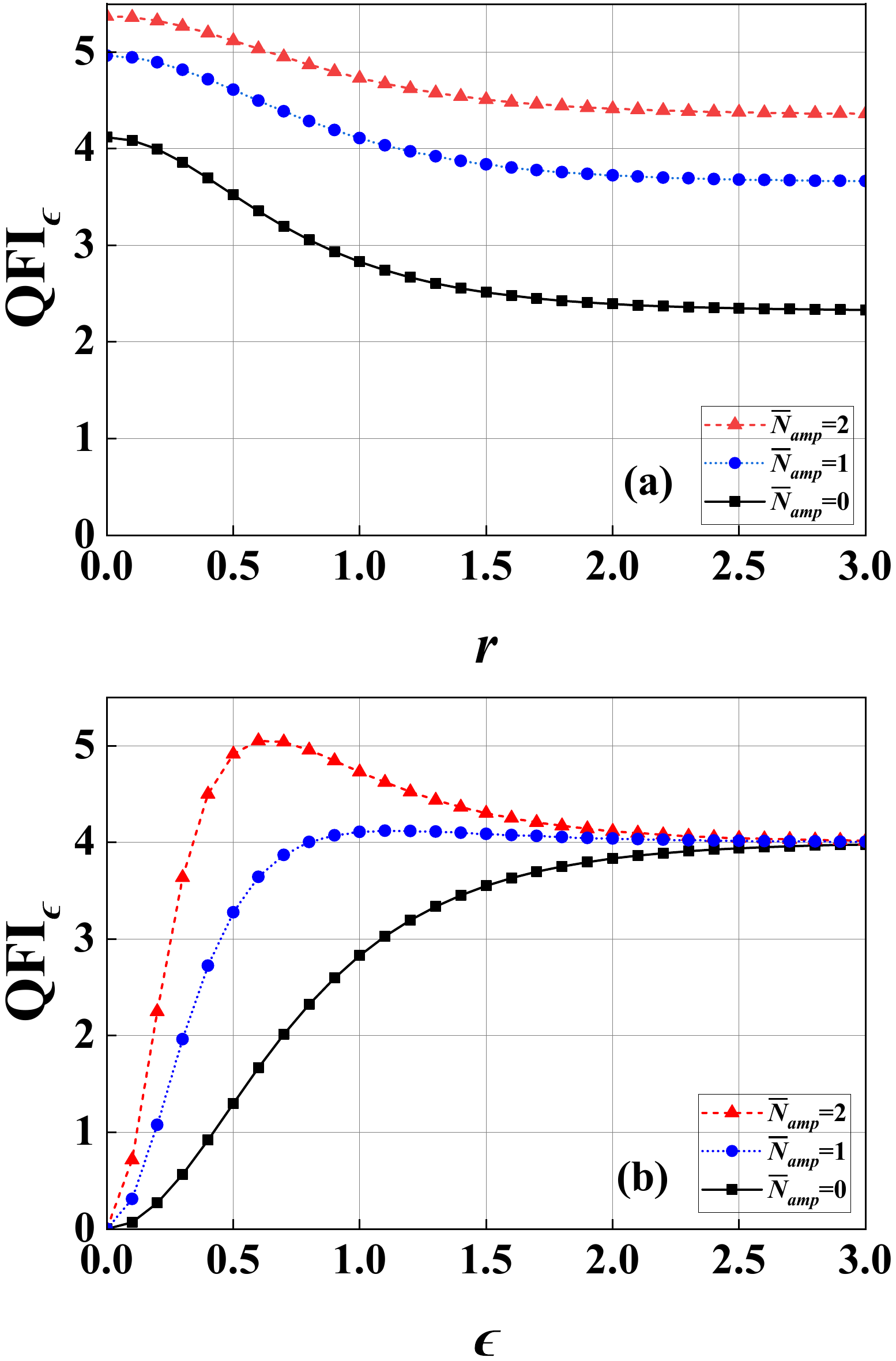}%
\newline
\caption{{}(Color online) The QFI with respect to the amplifier parameter $%
\protect\epsilon $ for the input squeezed vacuum state is analyzed as a
function of (a) the acceleration parameter $r$\ with $\protect\epsilon =1$
and\ $s=0.5,$ and of (b) the amplifier parameter $\protect\epsilon $\ with $%
r=1$ and\ $s=0.5.$}
\end{figure}
\begin{equation}
F_{\epsilon }^{\alpha }=\Xi _{1}\left( \frac{\Xi _{2}}{\Xi _{3}}+\frac{\Xi
_{4}}{\Xi _{5}+\Xi _{6}}\right) ,  \label{39}
\end{equation}%
where%
\begin{eqnarray}
\Xi _{1} &=&4\cosh ^{2}r,  \notag \\
\Xi _{2} &=&\left\vert \alpha \right\vert ^{2}\sinh ^{2}\epsilon ,  \notag \\
\Xi _{3} &=&\cosh ^{2}r[(1+\bar{N}_{\text{amp}})\cosh (2\epsilon )-\bar{N}_{%
\text{amp}}]  \notag \\
&&+\sinh ^{2}r,  \notag \\
\Xi _{4} &=&4(1+\bar{N}_{\text{amp}})^{2}\sinh ^{2}2\epsilon ,  \notag \\
\Xi _{5} &=&3\bar{N}_{\text{amp}}(2+\bar{N}_{\text{amp}})-4(1+\bar{N}_{\text{%
amp}})^{2}\cosh (2\epsilon )-5,  \notag \\
\Xi _{6} &=&2\cosh (2r)[1-\bar{N}_{\text{amp}}+(1+\bar{N}_{\text{amp}})\cosh
(2\epsilon )]^{2}  \notag \\
&&+(1+\bar{N}_{\text{amp}})^{2}\cosh (4\epsilon ).  \label{40}
\end{eqnarray}

Under the specific conditions where $\bar{N}_{\text{amp}}$ is equal to zero
and $r$ tends toward zero, Eq. (\ref{39}) converges to the results
established in the previous work \cite{28}. In Fig. 12(a), we show the QFI
of the amplifier parameter changing with the acceleration parameter $r.$ As
the acceleration parameter $r$ increases, the QFI exhibits a monotonic
decrease, asymptotically converging to a finite non-zero value in the limit
of infinite acceleration. This behavior is in complete agreement with the
findings reported in Ref. \cite{54}. Furthermore, we investigate the
influences of the amplifier parameter $\epsilon $ on the QFI, as depicted in
Fig. 12(b). It is clear that, in the rang of $\epsilon <1.12,$ the value of
the QFI for $\bar{N}_{\text{amp}}=0$ is lower than that for the cases of $%
\bar{N}_{\text{amp}}=1$ and $\bar{N}_{\text{amp}}=2,$ but the former is
larger than the latter when $\epsilon $ is greater than $1.12.$%
\begin{figure}[tbp]
\label{Fig14} \centering\includegraphics[width=0.8\columnwidth]{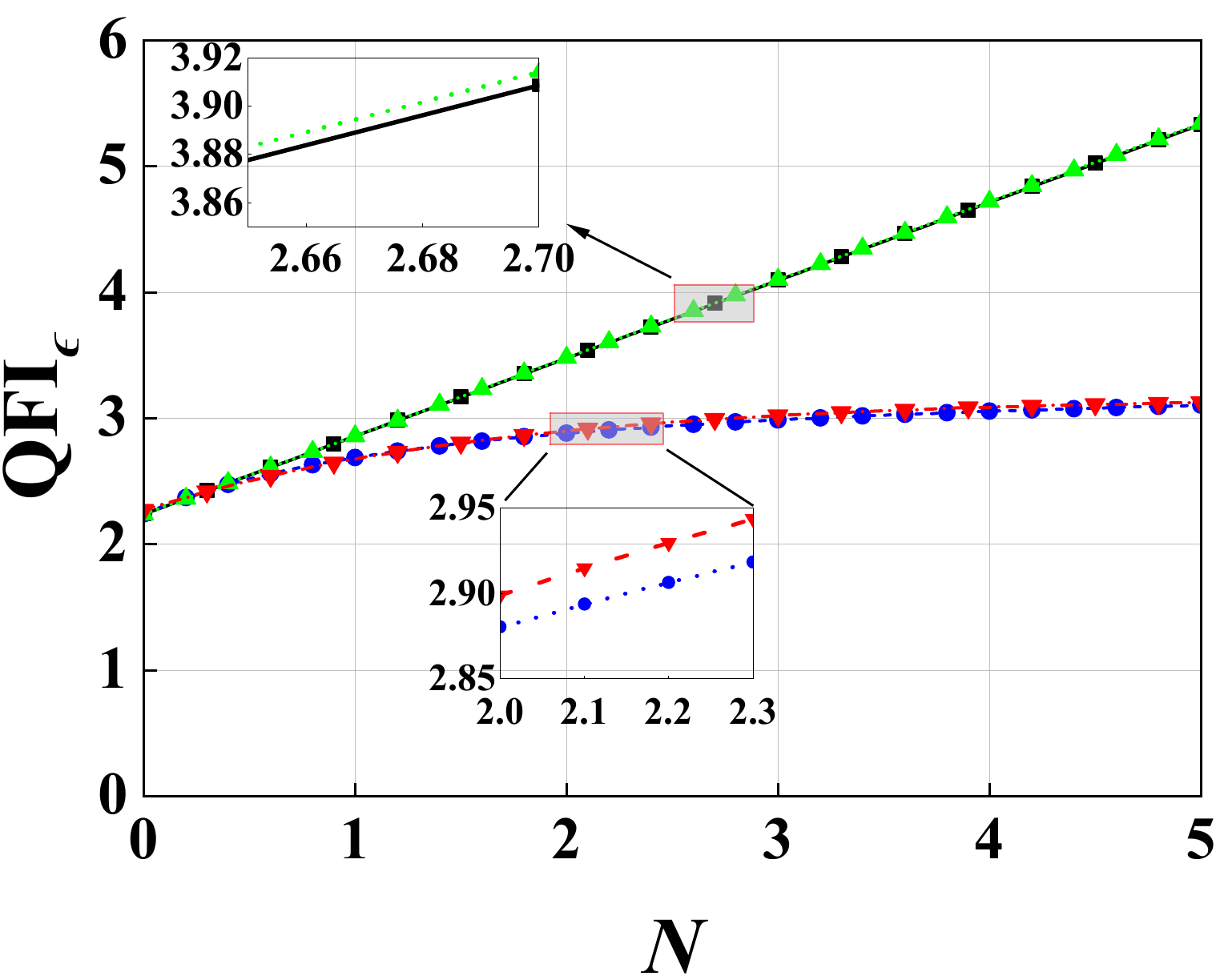}%
\newline
\caption{{}(Color online) The QFI with respect to the amplifier parameter $%
\protect\epsilon $\ is analyzed as a function of the mean photon number $N$\
\ with $\protect\epsilon =0.5,$\ $r=0.5$\ and\ $\bar{N}_{\text{amp}}=0.$\
The black and blue lines represent the true QFI for the input coherent state
and the squeezed vacuum state, respectively. The green and red lines depict
the numerical fitting results obtained from Eq. (\protect\ref{41}).}
\end{figure}

We further explore the QFI of the amplifier parameter $\epsilon $ by
utilizing a squeezed vacuum state as the input state. Owing to the
complexity of deriving analytical results, we concentrate on numerical
computations to evaluate the QFI changing with the relevant physical
parameters $r$ and $\epsilon $, as illustrated in Fig. 13. Similar to the
case of the input coherent state, in Fig. 13(a), we demonstrate that the QFI
exhibits a gradual decline as the acceleration parameter $r$ increases.
Furthermore, in Fig. 13(b), our results reveal distinct trends; while the
QFI values for both $\bar{N}_{\text{amp}}=0$ and $\bar{N}_{\text{amp}}=1$
configurations show an increase with the amplifier parameter $\epsilon $,
the QFI for $\bar{N}_{\text{amp}}=2$ displays a non-monotonic behavior,
initially increasing before reaching a maximum and subsequently decreasing
with increasing $\epsilon .$ For values of $\epsilon $ exceeding $2.5$, the
corresponding QFI for $\bar{N}_{\text{amp}}=0,$ $\bar{N}_{\text{amp}}=1,$
and $\bar{N}_{\text{amp}}=2$ approaches convergence, ultimately reaching a
nearly identical, non-zero asymptotic value. We further explore how the
choice of input state affects the QFI of the amplifier parameter $\epsilon $%
. As shown in Fig. 14, the QFI for both the coherent and squeezed vacuum
states grows monotonically with the mean photon number $N$. To quantify the
functional dependence and scaling behavior, the numerical results are fitted
to the following empirical expressions%
\begin{eqnarray}
F_{\epsilon }^{\alpha } &\simeq &2.24+0.62N,  \notag \\
F_{\epsilon }^{s} &\simeq &3.17-0.9e^{-0.6N},  \label{41}
\end{eqnarray}%
where $F_{\epsilon }^{\alpha }$\ and $F_{\epsilon }^{s}$\ represent the QFI
of the amplifier parameter for the coherent state and the squeezed vacuum
state, respectively. As shown in Fig. 14, these empirical expressions can
saturate the corresponding true QFI.\textbf{\ } 
\begin{figure}[tbp]
\label{Fig15} \centering\includegraphics[width=0.8\columnwidth]{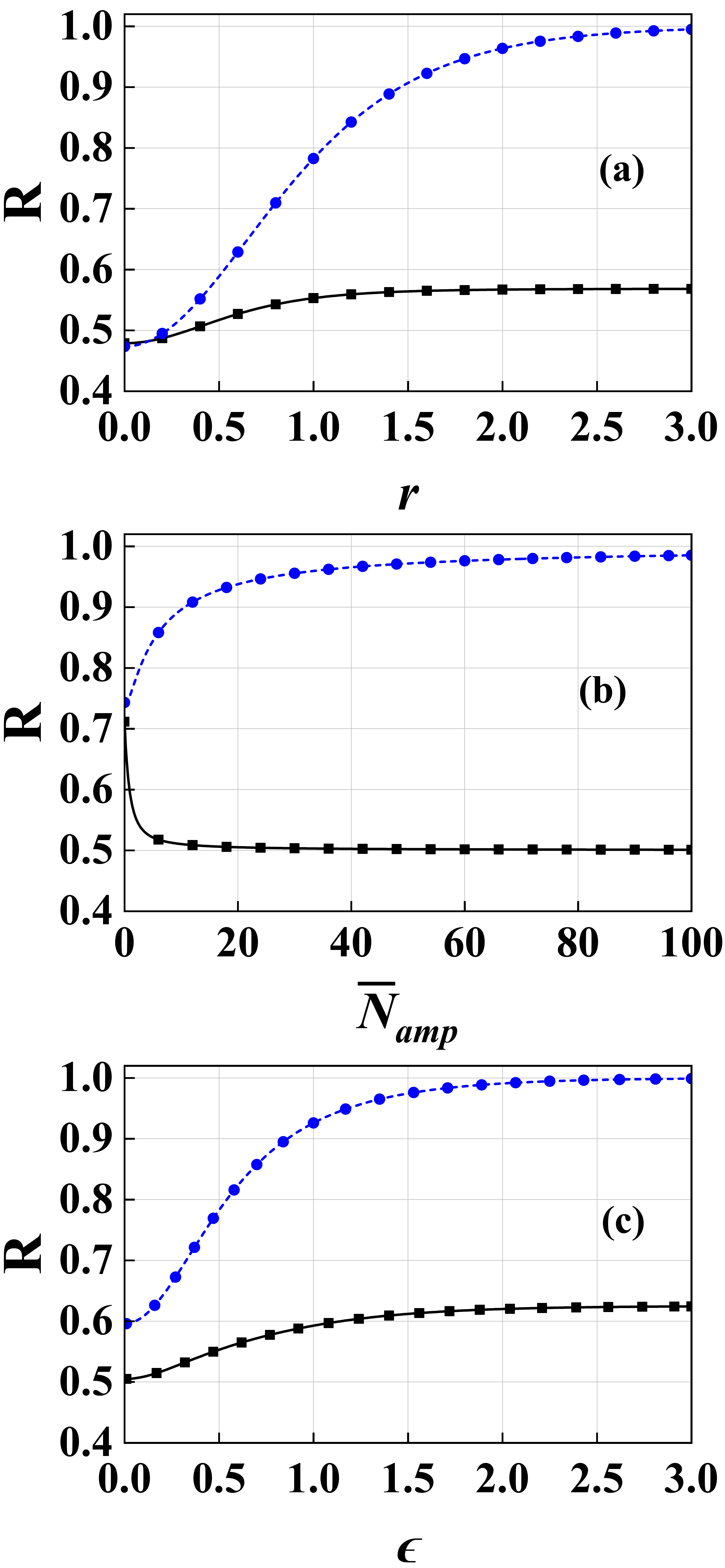}%
\newline
\caption{{}(Color online) The ratio $R$\ between the CFI and the QFI with
respect to the amplifier parameter $\protect\epsilon $\ for the input
coherent state is analyzed as a function of (a) the acceleration parameter $%
r $\ with $\protect\epsilon =0.5,$\ Re$(\protect\alpha )^{2}=2,$\ Im$(%
\protect\alpha )=0,$\ and $\bar{N}_{\text{amp}}=2,$\ of (b) the thermal mean
number $\bar{N}_{\text{amp}}$\ with $\protect\epsilon =0.5,$\ $r=1,$\ Re$(%
\protect\alpha )^{2}=2,$\ and Im$(\protect\alpha )=0,$\ and of (c) the
amplifier parameter $\protect\epsilon $\ with $r=1,$\ Re$(\protect\alpha %
)^{2}=2,$\ Im$(\protect\alpha )=0,$\ and $\bar{N}_{\text{amp}}=2.$\ The
black and blue lines correspond to homodyne and heterodyne measurements,
respectively. }
\end{figure}

We now investigate the conditions required for the CFI for Gaussian
measurements, including homodyne and heterodyne measurements, to saturate
the QFI of the amplifier parameter. Following Eq. (\ref{12}), the analytical
expressions for the CFI with a coherent state input under both measurements
schemes are derived as follows:%
\begin{eqnarray}
\tilde{F}_{\epsilon }^{\text{hom}} &=&\frac{\Xi _{7}\Xi _{8}+\Xi _{9}}{\Xi
_{3}^{2}},  \notag \\
\tilde{F}_{\epsilon }^{\text{het}} &=&(\Xi _{10}+\Xi _{11})\Xi _{12},
\label{42}
\end{eqnarray}%
where $\tilde{F}_{\epsilon }^{\text{hom}}$\ and $\tilde{F}_{\epsilon }^{%
\text{het}}$\ denote the CFI for homodyne and heterodyne measurements,
respectively, $\Xi _{3}$\ is given by Eq. (\ref{40}) and%
\begin{eqnarray}
\Xi _{7} &=&(1+\bar{N}_{\text{amp}})(1+\bar{N}_{\text{amp}}+\text{Re}(\alpha
)^{2})\cosh (2\epsilon )  \notag \\
&&+(1+\bar{N}_{\text{amp}})^{2}-\bar{N}_{\text{amp}}\text{Re}(\alpha )^{2}, 
\notag \\
\Xi _{8} &=&4\cosh ^{4}r\sinh ^{2}\epsilon ,  \notag \\
\Xi _{9} &=&4\text{Re}(\alpha )^{2}\sinh ^{2}r\cosh ^{2}r\sinh ^{2}\epsilon ,
\notag \\
\Xi _{10} &=&2(1+\bar{N}_{\text{amp}})^{2}+(1-\bar{N}_{\text{amp}%
})\left\vert \alpha \right\vert ^{2},  \notag \\
\Xi _{11} &=&(1+\bar{N}_{\text{amp}})(\left\vert \alpha \right\vert ^{2}+2+2%
\bar{N}_{\text{amp}})\cosh (2\epsilon ),  \notag \\
\Xi _{12} &=&\frac{4\sinh ^{2}\epsilon }{(1-\bar{N}_{\text{amp}}+(1+\bar{N}_{%
\text{amp}})\cosh (2\epsilon ))^{2}}.  \label{43}
\end{eqnarray}%
\begin{figure}[tbp]
\label{Fig16} \centering\includegraphics[width=0.8\columnwidth]{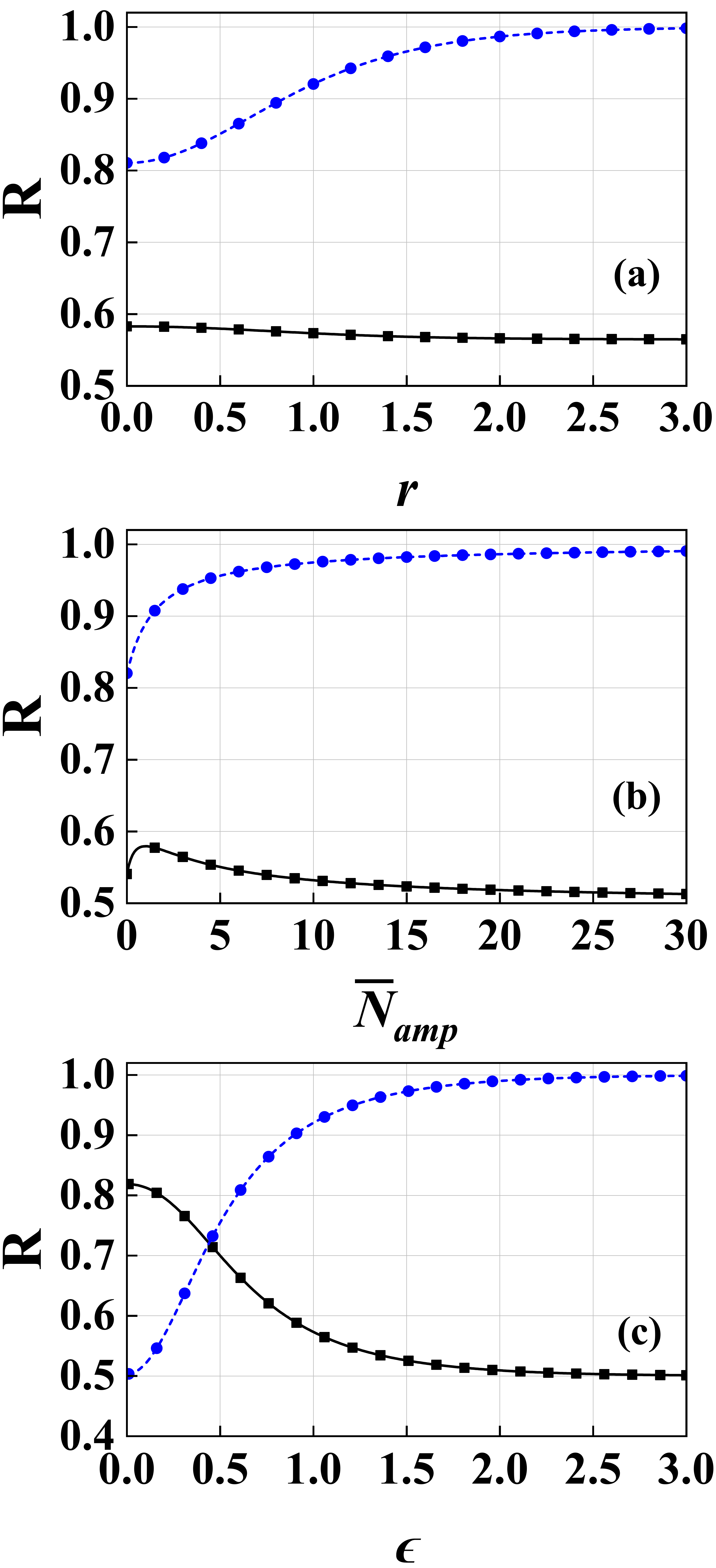}%
\newline
\caption{{}(Color online) The ratio $R$\ between the CFI and the QFI with
respect to the amplifier parameter $\protect\epsilon $\ for the input
squeezed vacuum state is analyzed as a function of (a) the acceleration
parameter $r$\ with $\protect\epsilon =1,$\ $s=0.5,$\ and $\bar{N}_{\text{amp%
}}=2,$\ of (b) the thermal mean number $\bar{N}_{\text{amp}}$\ with $\protect%
\epsilon =1,$\ $r=1,$\ $s=0.5,$\ and of (c) the amplifier parameter $\protect%
\epsilon $\ with $r=1,$\ $s=0.5,$\ and $\bar{N}_{\text{amp}}=2.$\ The black
and blue lines correspond to homodyne and heterodyne measurements,
respectively. }
\end{figure}
Obviously, the CFI $\tilde{F}_{\epsilon }^{\text{het}}$\ for heterodyne
measurement is found to approach the QFI $F_{\epsilon }^{\alpha }$\ in the
high-acceleration limit, i.e., $\tilde{F}_{\epsilon }^{\text{het}%
}=\lim_{r\rightarrow \infty }F_{\epsilon }^{\alpha }.$ In the large-$\bar{N}%
_{\text{amp}}$\ limit $\bar{N}_{\text{amp}}\rightarrow \infty ,$\ we obatin
the relation $F_{\epsilon }^{\alpha }=\tilde{F}_{\epsilon }^{\text{het}}=2%
\tilde{F}_{\epsilon }^{\text{hom}}=4\coth ^{2}\epsilon .$ Similarly, in the
large-$\epsilon $\ limit $\epsilon \rightarrow \infty ,$ we find $%
F_{\epsilon }^{\alpha }=\tilde{F}_{\epsilon }^{\text{het}}=4+\frac{%
2\left\vert \alpha \right\vert ^{2}}{1+\bar{N}_{\text{amp}}}$ and $\tilde{F}%
_{\epsilon }^{\text{hom}}=2+\frac{2\text{Re}(\alpha )^{2}}{1+\bar{N}_{\text{%
amp}}}.$ To visualize these relationships between the CFI and the QFI, we
plot the ratio $R$\ between the CFI and the QFI\ as a function of the
relevant physical parameters $r,$\ $\bar{N}_{\text{amp}},$\ and $\varepsilon
,$\ as shown in Fig. 15. In Fig. 15(a), the ratio for heterodyne measurement
exhibits a continuous and monotonic increase starting from approximately $%
0.48$. As the acceleration parameter $r$\ grows, the ratio climbs
significantly and asymptotically approaches the theoretical optimum value $1$%
\ for $r\geq 2.5$. This trend confirms that heterodyne measurement can
saturate the QFI in the high-acceleration regime. In contrast, homodyne
measurement shows a much more constrained performance. Although the ratio
initially increases with $r$, it quickly reaches a plateau, saturating at a
value of approximately $0.57$. In Fig. 15(b), the ratio $R$\ for heterodyne
measurement shows a rapid and monotonic increase in the low-$\bar{N}_{\text{%
amp}}$\ regime. Starting from approximately $0.74$, the ratio quickly
surpasses $0.9$\ and asymptotically approaches the theoretical optimum value 
$1$\ as $\bar{N}_{\text{amp}}$\ increases. In sharp contrast, the
performance of homodyne measurement exhibits a significant decline. The
ratio drops abruptly from its initial value and swiftly stabilizes at a
plateau of $0.5$\ for $\bar{N}_{\text{amp}}\geq 10$. In Fig. 15(c), the
ratio $R$\ for heterodyne measurement exhibits a monotonic and significant
increase as $\epsilon $\ grows. Starting from a sub-optimal value of
approximately $0.6$, the curve climbs steeply and asymptotically approaches
the theoretical maximum value $1$. This demonstrates that in the
strong-amplification regime, heterodyne measurement becomes increasingly
efficient, eventually reaching the QFI. In contrast, homodyne measurement
shows a much more limited performance gain. While the ratio for homodyne
measurement also increases with $\epsilon $, it does so much more gradually
and quickly reaches a plateau at a significantly lower value of
approximately $0.62$. 
\begin{figure}[tbp]
\label{Fig17} \centering\includegraphics[width=0.8\columnwidth]{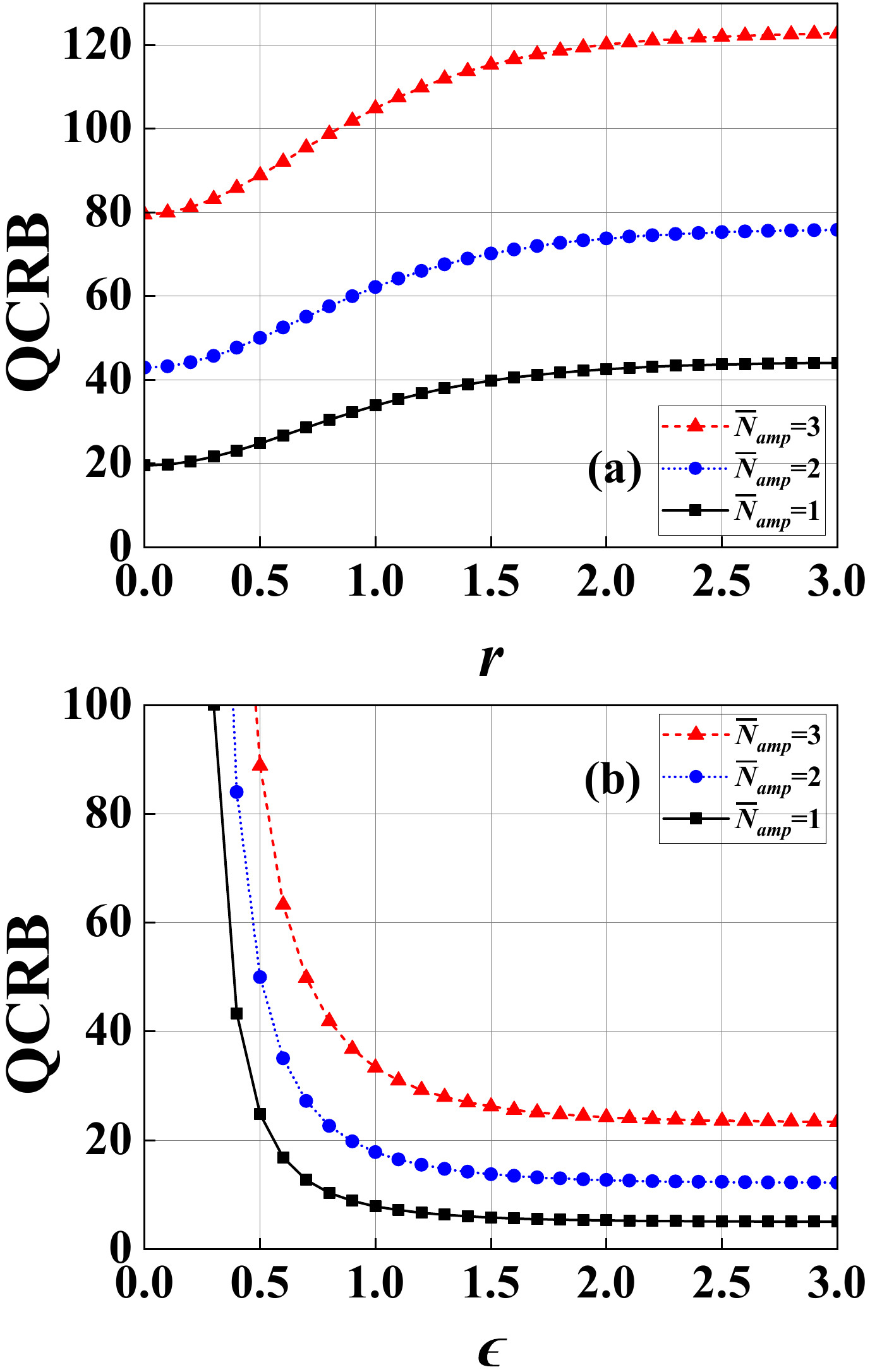}%
\newline
\caption{{}(Color online) The QCRB with respect to the two-parameter
estimation including the amplifier parameter $\protect\epsilon $\ and the
thermal mean number $\bar{N}_{\text{amp}}$\ of the thermal amplifier channel
for the input coherent state is analyzed as a function of (a) the
acceleration parameter $r$\ with $\protect\epsilon =0.5$ and\ $\left\vert 
\protect\alpha \right\vert ^{2}=18,$ and of (b) the amplifier parameter $%
\protect\epsilon $\ with $r=0.5$ and\ $\left\vert \protect\alpha \right\vert
^{2}=18.$}
\end{figure}
\begin{figure}[tbp]
\label{Fig18} \centering\includegraphics[width=0.8\columnwidth]{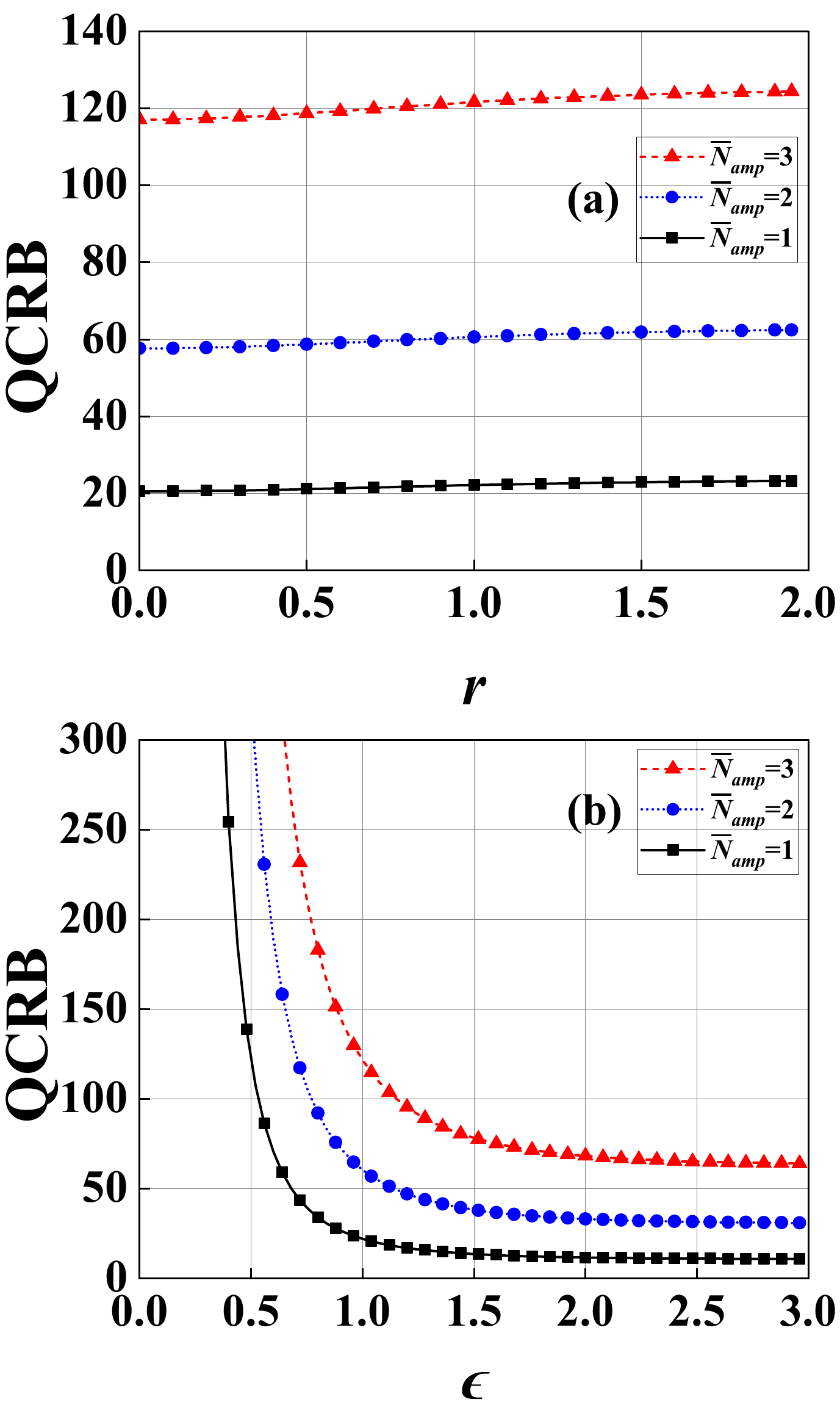}%
\newline
\caption{{}(Color online) The QCRB with respect to the two-parameter
estimation including the amplifier parameter $\protect\epsilon $\ and the
thermal mean number $\bar{N}_{\text{amp}}$\ of the thermal amplifier channel
for the input squeezed vacuum state is analyzed as a function of (a) the
acceleration parameter $r$\ with $\protect\epsilon =1$ and\ $s=2,$ and of
(b) the amplifier parameter $\protect\epsilon $\ with $r=1$ and\ $s=2.$}
\end{figure}

Regarding the input squeezed vacuum state, we perform numerical computations
to investigate the evolution of the ratio $R$\ as a function of the relevant
physical parameters $r$, $\bar{N}_{\text{amp}}$, and $\epsilon $, as
illustrated in Fig. 16. In Fig. 16(a), the ratio for heterodyne measurement
starts at a relatively high initial value of approximately $0.8$\ at the
inertial limit ($r\rightarrow 0$). It then exhibits a monotonic increase as $%
r$\ grows, eventually approaching the theoretical optimum of unity for $%
r\geq 2.5$. This indicates that heterodyne measurement is highly efficient
across the entire acceleration range and saturates the QFI in the
high-acceleration regime. In contrast, homodyne measurement shows a slightly
declining trend. The ratio starts near $0.58$\ and decreases marginally as $%
r $\ increases, eventually stabilizing at a value of approximately $0.56$.
The increasing divergence between the two curves further emphasizes that
heterodyne measurement is far superior for maintaining estimation precision
under non-inertial effects, whereas homodyne measurement fails to extract
the additional quantum information available at higher accelerations. In
Fig. 16(b), the ratio for heterodyne measurement exhibits a rapid and
monotonic increase with $\bar{N}_{\text{amp}}$. Starting from an initial
value near $0.82$, it swiftly approaches the theoretical maximum of unity.
This behavior indicates that heterodyne measurement can saturate the QCRB in
the large-$\bar{N}_{\text{amp}}$\ limit. In contrast, homodyne measurement
follows a non-monotonic trend. The ratio initially shows a slight increase,
reaching a local peak of approximately $0.58$\ at small $\bar{N}_{\text{amp}%
} $, before gradually declining and asymptotically approaching $0.5$\ as the
thermal mean number increases. In Fig. 16(c), the ratio for heterodyne
measurement starts at a lower bound of $0.5$\ and exhibits a monotonic
increase as the amplifier parameter grows. It rapidly climbs and
asymptotically approaches the theoretical optimum of unity, confirming its
ability to saturate the QFI in the strong-amplification regime. In contrast,
homodyne measurement shows the opposite behavior. The ratio begins at a
higher value of approximately $0.82$\ in the limit of small $\epsilon $, but
decreases significantly as $\epsilon $\ increases. It eventually plateaus at
a value of $0.5$\ in the large-$\epsilon $\ limit. The intersection of the
two curves around $\epsilon \approx 0.45$\ marks the point where heterodyne
measurement begins to outperform homodyne measurement, highlighting that
while homodyne measurement may be preferable for weak amplification,
heterodyne measurement is far more effective for characterizing channels
with high gain.\textbf{\ } 
\begin{figure}[tbp]
\label{Fig19} \centering\includegraphics[width=0.8\columnwidth]{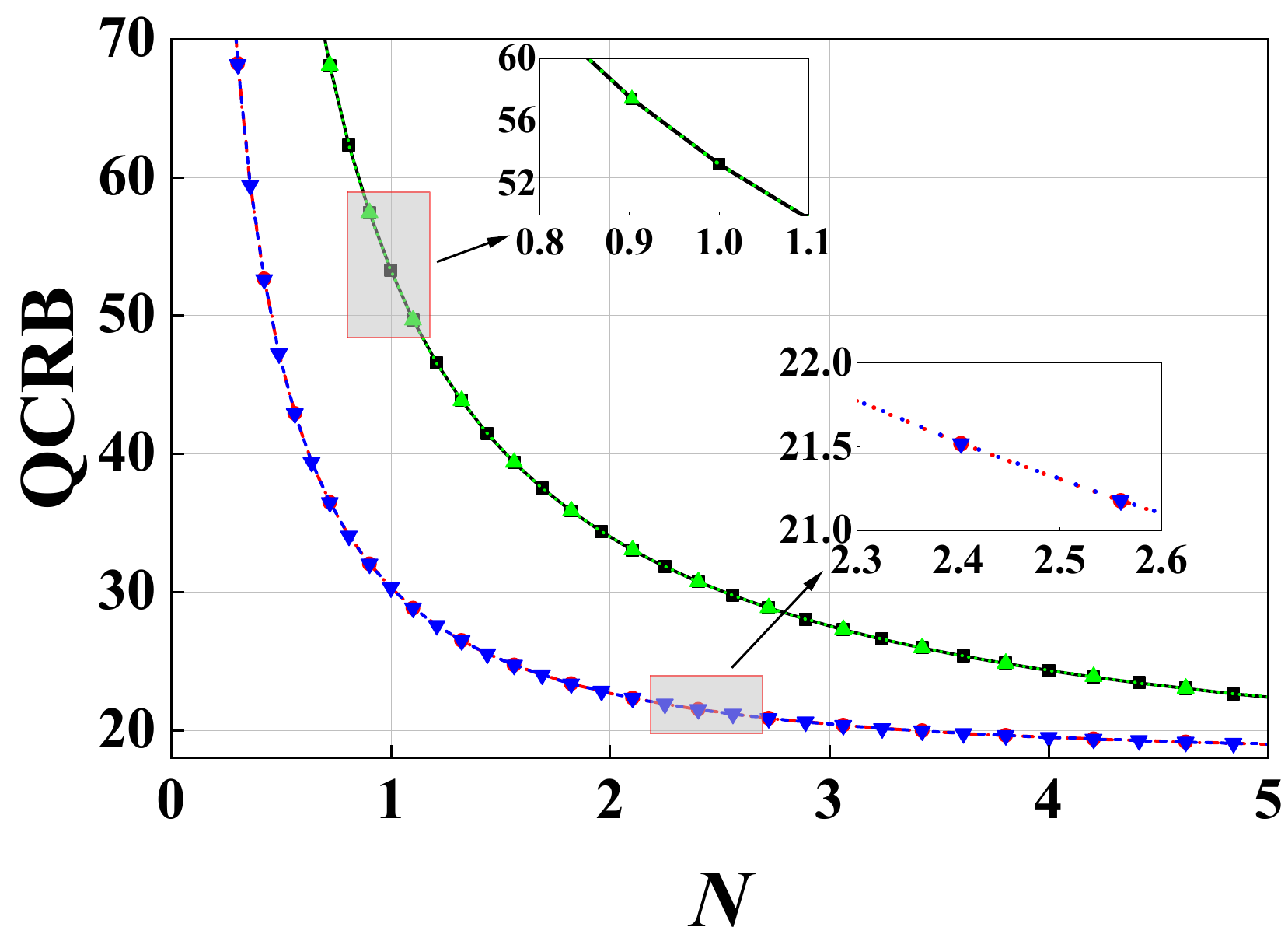}%
\newline
\caption{{}(Color online) The QCRB with respect to the two-parameter
estimation including the amplifier parameter $\protect\epsilon $\ and the
thermal mean number $\bar{N}_{\text{amp}}$\ of the thermal amplifier channel
for the input squeezed vacuum state is analyzed as a function of (a) the
acceleration parameter $r$\ with $\protect\epsilon =1$\ and\ $s=2,$\ and of
(b) the amplifier parameter $\protect\epsilon $\ with $r=1$\ and\ $s=2.$\
The black and blue lines represent the true QCRB for the input coherent
state and the squeezed vacuum state, respectively. The green and red lines
depict the numerical fitting results obtained from Eq. (\protect\ref{47}).}
\end{figure}

Subsequently, we delve into the analysis of a two-parameter estimation
scenario, which encompasses the amplifier parameter $\epsilon $ and the
thermal mean number $\bar{N}_{\text{amp}}$ of the thermal amplifier channel.
For an input coherent state, as delineated by Eqs. (\ref{19}) and (\ref{38}%
), the commutation relation between the symmetric logarithmic derivative
operators $\hat{L}_{\epsilon }$ and $\hat{L}_{\bar{N}_{\text{amp}}}$ can be
articulated as 
\begin{equation}
\lbrack \hat{L}_{\epsilon },\hat{L}_{\bar{N}_{\text{amp}}}]=2i\left[ (\Phi
_{1}+\Phi _{3})\hat{x}_{1}+(\Phi _{2}+\Phi _{4})\hat{p}_{1}\right] ,
\label{44}
\end{equation}%
where $\Phi _{1},$ $\Phi _{2},$ $\Phi _{3},$ and $\Phi _{4}$ are given in
Appendix B, not shown here for simplicity. It is evident that the symmetric
logarithmic derivative operators $\hat{L}_{\epsilon }$ and $\hat{L}_{\bar{N}%
_{\text{amp}}}$ are non-commutative. However, a noteworthy observation is
that the corresponding mean Uhlmann curvature matrix is found to be a zero
matrix. This suggests that the QCRB acts as an asymptotically tight
precision limit. In this context, we derive the QCRB with respect to the
two-parameter estimation including the amplifier parameter $\epsilon $ and
the thermal mean number $\bar{N}_{\text{amp}}$ of the thermal amplifier
channel for a coherent state input%
\begin{equation}
Q_{\text{amp}}^{\alpha }=\lambda _{1}[\lambda _{2}(\lambda _{3}+\lambda
_{4})-\lambda _{5}],  \label{45}
\end{equation}%
where%
\begin{eqnarray}
\lambda _{1} &=&\frac{1}{8\left\vert \alpha \right\vert ^{2}}\text{csch}%
^{4}\epsilon ,  \notag \\
\lambda _{2} &=&1-\bar{N}_{\text{amp}}+(1+\bar{N}_{\text{amp}})\cosh
(2\epsilon ),  \notag \\
\lambda _{3} &=&3-2\left\vert \alpha \right\vert ^{2}\text{(}\bar{N}_{\text{%
amp}}-1\text{)}+8\bar{N}_{\text{amp}}+4\bar{N}_{\text{amp}}^{2},  \notag \\
\lambda _{4} &=&\cosh (2\epsilon )[5+8\bar{N}_{\text{amp}}+4\bar{N}_{\text{%
amp}}^{2}  \notag \\
&&+2\left\vert \alpha \right\vert ^{2}(1+\bar{N}_{\text{amp}})],  \notag \\
\lambda _{5} &=&(\lambda _{6}+\lambda _{7})\text{sech}^{2}r,  \notag \\
\lambda _{6} &=&3+4\left\vert \alpha \right\vert ^{2}+4\bar{N}_{\text{amp}%
}(2+\bar{N}_{\text{amp}}-\left\vert \alpha \right\vert ^{2}),  \notag \\
\lambda _{7} &=&\cosh (2\epsilon )[5+4\left\vert \alpha \right\vert ^{2} 
\notag \\
&&+4\bar{N}_{\text{amp}}(2+\bar{N}_{\text{amp}}+\left\vert \alpha
\right\vert ^{2})].  \label{46}
\end{eqnarray}

To elucidate the impact of the Unruh effect on the QCRB with respect to the
two-parameter estimation including the amplifier parameter and the thermal
mean number of the thermal amplifier channel, we depict the QCRB as a
function of the acceleration parameter $r$ in Fig. 17(a). As evident from
the plot, the QCRB exhibits a monotonic increase with rising values of $r$,
ultimately approaching a finite, non-zero asymptotic value in the limit of
infinite acceleration. Furthermore, in Fig. 17(b), we demonstrate the
impacts of the amplifier parameter $\epsilon $ on the QCRB, revealing a
gradual decrease in the QCRB as $\epsilon $ increases, which similarly
converges to a non-zero asymptotic value. 
\begin{figure}[tbp]
\label{Fig20} \centering\includegraphics[width=0.8\columnwidth]{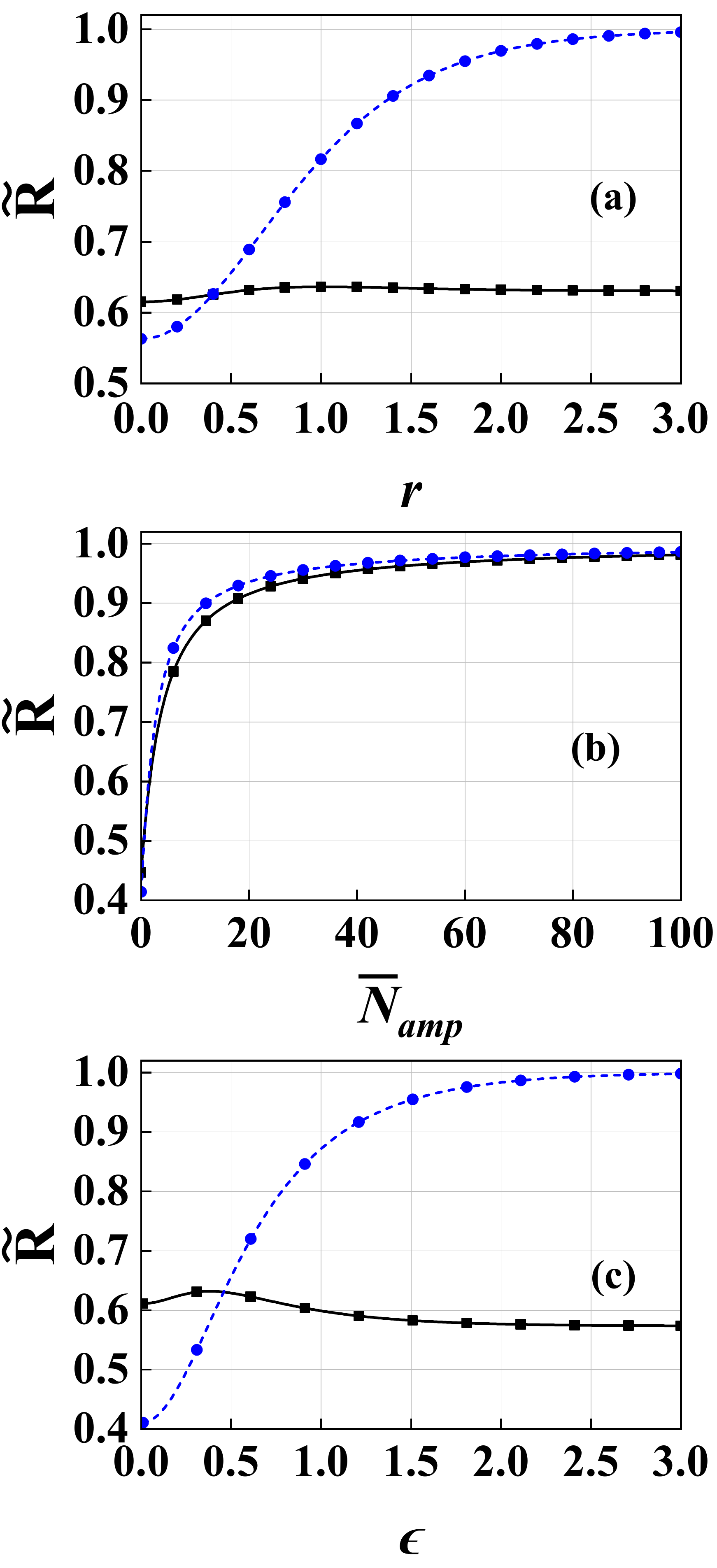}%
\newline
\caption{{}(Color online) The ratio $\tilde{R}$\ between the QCRB and the
CRB with respect to the two-parameter estimation including the amplifier
parameter $\protect\epsilon $\ and the thermal mean number $\bar{N}_{\text{%
amp}}$\ of the thermal amplifier channel for the input coherent state is
analyzed as a function of (a) the acceleration parameter $r$\ with $\protect%
\epsilon =0.5,$\ Re$(\protect\alpha )^{2}=18,$\ Im$(\protect\alpha )=0,$\
and $\bar{N}_{\text{amp}}=2,$\ of (b) the thermal mean number $\bar{N}_{%
\text{amp}}$\ with $\protect\epsilon =0.5,$\ $r=0.5,$\ Re$(\protect\alpha %
)^{2}=18,$\ and Im$(\protect\alpha )=0,$\ and of (c) the amplifier parameter 
$\protect\epsilon $\ with $r=0.5,$\ Re$(\protect\alpha )^{2}=18,$\ Im$(%
\protect\alpha )=0,$\ and $\bar{N}_{\text{amp}}=2.$\ The black and blue
lines correspond to homodyne and heterodyne measurements, respectively. }
\end{figure}
\begin{figure}[tbp]
\label{Fig21} \centering\includegraphics[width=0.8\columnwidth]{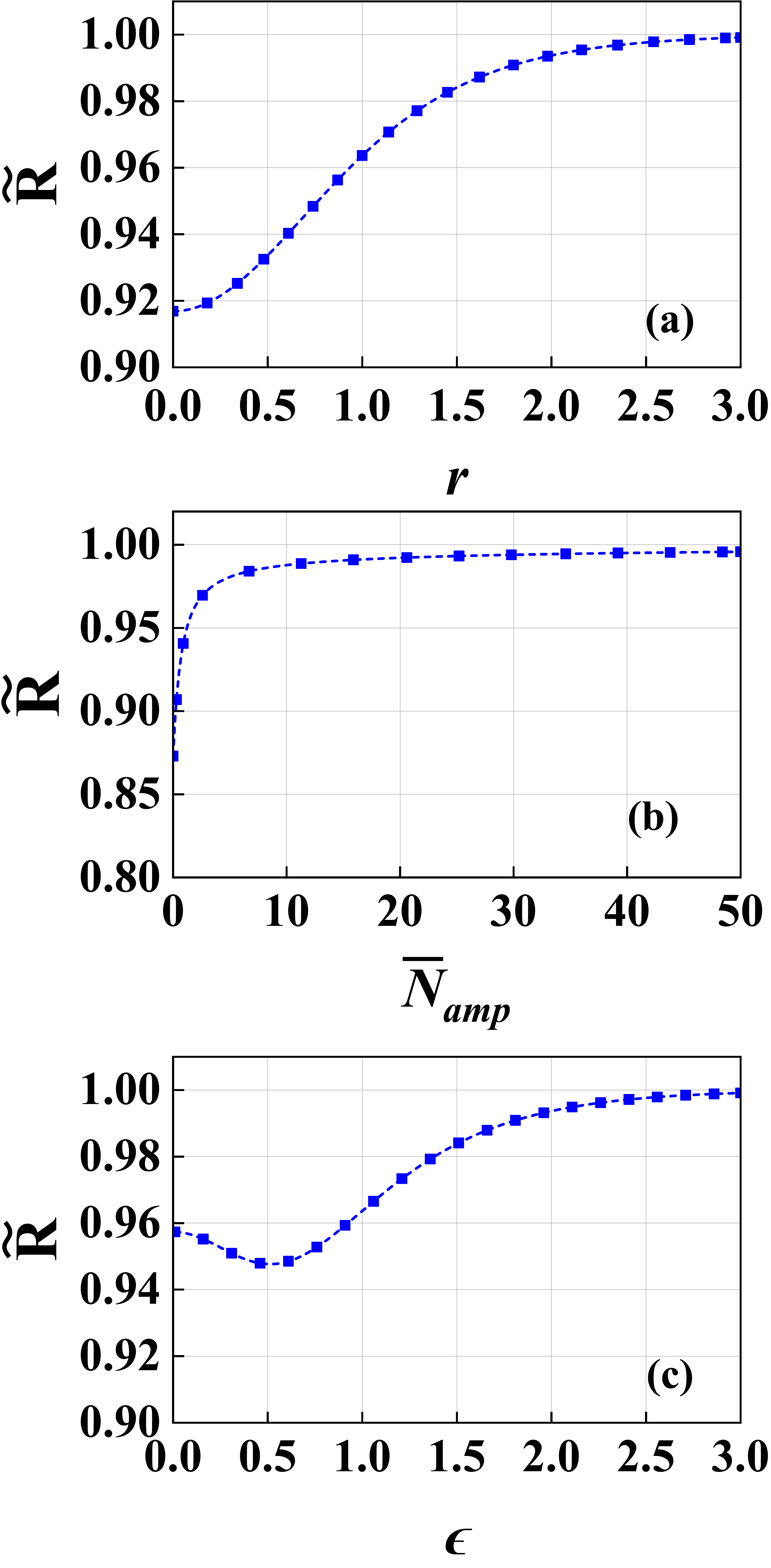}%
\newline
\caption{{}(Color online) The ratio $\tilde{R}$\ between the QCRB and the
CRB with respect to the two-parameter estimation including the amplifier
parameter $\protect\epsilon $\ and the thermal mean number $\bar{N}_{\text{%
amp}}$\ of the thermal amplifier channel for the input squeezed vacuum state
is analyzed as a function of (a) the acceleration parameter $r$\ with $%
\protect\epsilon =1,$\ $s=2,$\ and $\bar{N}_{\text{amp}}=2,$\ of (b) the
thermal mean number $\bar{N}_{\text{amp}}$\ with $\protect\epsilon =1,$\ $%
r=1,$\ $s=2,$\ and of (c) the amplifier parameter $\protect\epsilon $\ with $%
r=1,$\ $s=2,$\ and $\bar{N}_{\text{amp}}=2.$\ The homodyne measurement fails
to simultaneously estimate $\protect\epsilon $\ and $\bar{N}_{\text{amp}}$\
due to the singularity of the classical Fisher information matrix, which
results in a divergent CRB. Consequently, we exclude the homodyne
measurement results and focus our discussion on the performance of the
heterodyne measurement. }
\end{figure}

For the input squeezed vacuum state, we can also find that the symmetric
logarithmic derivative operators $\hat{L}_{\epsilon }$ and $\hat{L}_{\bar{N}%
_{\text{amp}}}$ are also non-commutative. Despite this non-commutativity,
the associated mean Uhlmann curvature matrix is identified as a zero matrix,
signifying that the QCRB with respect to the two-parameter estimation
including the amplifier parameter and the thermal mean number of the thermal
amplifier channel acts as an asymptotically tight precision limit. Given the
intricate nature of obtaining analytical solutions, our attention is
directed toward numerical evaluations of the QCRB in relation to the
pertinent physical parameters, as depicted in Fig. 18. In line with the
situation where coherent states function as initial conditions, the QCRB
steadily escalates with higher values of $r$ and diminishes with larger $%
\epsilon $. Furthermore, we explore how the choice of input state affects
the ultimate precision limit. Fig. 19 depicts the evolution of the QCRB
against the mean photon number $N$\ for both the coherent state and the
squeezed vacuum state. In both scenarios, the QCRB diminishes monotonically
as $N$\ grows. To further describe the scaling behavior of these precision
limits, we perform a numerical fit on the QCRB, resulting in the following
empirical formulas%
\begin{eqnarray}
Q_{\text{amp}}^{\alpha } &\simeq &14.69+\frac{38.57}{N},  \notag \\
Q_{\text{amp}}^{s} &\simeq &18.22+\frac{16.70}{N},  \label{47}
\end{eqnarray}%
where $Q_{\text{amp}}^{\alpha }$\ and $Q_{\text{amp}}^{s}$\ denote the QCRB
for the coherent state and the squeezed vacuum state, respectively. As
illustrated in Fig. 19, these empirical formulas provide a precise fit that
saturates the corresponding true QCRB.

Subsequently, we explore the potential for Gaussian measurements, such as
homodyne and heterodyne measurements, to attain the ultimate precision limit
defined by the QCRB. Based on Eq. (\ref{12}), the corresponding CRB with
respect to the two-parameter estimation including the amplifier parameter $%
\epsilon $\ and the thermal mean number $\bar{N}_{\text{amp}}$\ for a
coherent state input are determined as%
\begin{eqnarray}
C_{\text{amp}}^{\text{hom}} &=&(\lambda _{8}+\lambda _{9}\lambda
_{10})\lambda _{11},  \notag \\
C_{\text{amp}}^{\text{het}} &=&(\lambda _{12}+\lambda _{13})\lambda _{14}
\label{48}
\end{eqnarray}%
where%
\begin{eqnarray}
\lambda _{8} &=&4\bar{N}_{\text{amp}}(2+\bar{N}_{\text{amp}}+\text{Re}%
(\alpha )^{2})  \notag \\
&&+5+4\text{Re}(\alpha )^{2},  \notag \\
\lambda _{9} &=&((1+\bar{N}_{\text{amp}})^{2}+\text{Re}(\alpha )^{2})\cosh
(2r)  \notag \\
&&+(1+\bar{N}_{\text{amp}})^{2},  \notag \\
\lambda _{10} &=&2\text{csch}^{2}\epsilon \text{sech}^{2}r,  \notag \\
\lambda _{11} &=&\frac{2+2\bar{N}_{\text{amp}}+(1+\tanh ^{2}r)\text{csch}%
^{2}\epsilon }{4\text{Re}(\alpha )^{2}},  \notag \\
\lambda _{12} &=&5+8\bar{N}_{\text{amp}}+4\bar{N}_{\text{amp}}^{2}+2(1+\bar{N%
}_{\text{amp}})\left\vert \alpha \right\vert ^{2},  \notag \\
\lambda _{13} &=&2(\left\vert \alpha \right\vert ^{2}+2(1+\bar{N}_{\text{amp}%
})^{2})\text{csch}^{2}\epsilon ,  \notag \\
\lambda _{14} &=&\frac{1+\bar{N}_{\text{amp}}+\text{csch}^{2}\epsilon }{%
2\left\vert \alpha \right\vert ^{2}}.  \label{49}
\end{eqnarray}%
Specifically, the CRB $C_{\text{amp}}^{\text{het}}$\ for heterodyne
measurement is found to asymptotically coincide with the QCRB in the
high-acceleration limit, i.e., $C_{\text{amp}}^{\text{het}}=\lim_{r\mapsto
\infty }Q_{\text{amp}}^{\alpha }.$ Similarly, in the large-$\epsilon $\
limit $\epsilon \rightarrow \infty ,$ we find that 
\begin{eqnarray}
&&Q_{\text{amp}}^{\alpha }  \notag \\
&=&C_{\text{amp}}^{\text{het}}  \notag \\
&=&\frac{(1+\bar{N}_{\text{amp}})(5+4\bar{N}_{\text{amp}}(2+\bar{N}_{\text{%
amp}}))}{2\left\vert \alpha \right\vert ^{2}}  \notag \\
&&+(1+\bar{N}_{\text{amp}})^{2}.  \label{50}
\end{eqnarray}%
To illustrate the relationships between the CRB and the QCRB, the ratio $%
\tilde{R}$\ between the QCRB and the CRB is plotted against the relevant
physical parameters $r,$\ $\bar{N}_{\text{amp}},$\ and $\epsilon ,$\ as
presented in Fig. 20. In Fig. 20(a), the ratio for heterodyne measurement $%
\tilde{R}$\ begins at approximately $0.56$\ in the inertial limit ($%
r\longrightarrow 0$). As the acceleration parameter $r$\ increases, the
ratio undergoes a significant and monotonic rise, asymptotically approaching
the theoretical maximum of unity. This trend demonstrates that heterodyne
measurement effectively captures the full quantum information and saturates
the QCRB as the non-inertial effects become more dominant. Conversely,
homodyne measurement exhibits a nearly invariant performance across the
entire range of $r$. The ratio $\tilde{R}$\ stays localized around a much
lower value of approximately $0.63$, showing only a slight initial increase
before stabilizing into a plateau. The distinct crossover point near $%
r\approx 0.4$\ indicates that while homodyne measurement may offer a
marginal advantage at very low accelerations, heterodyne measurement becomes
the vastly superior strategy as $r$\ grows. In Fig. 20(b), the ratio $\tilde{%
R}$\ for heterodyne measurement starts from a relatively low value of
approximately $0.42$\ and climbs rapidly, asymptotically approaching the
theoretical maximum of unity. Homodyne measurement follows a nearly
identical trajectory, starting slightly higher at approximately $0.45$\ and
maintaining a close proximity to the heterodyne curve throughout the entire
range of $\bar{N}_{\text{amp}}$. As the thermal mean number increases, the
ratio for homodyne measurement also approaches unity, though it remains
marginally below the heterodyne curve in the large-$\bar{N}_{\text{amp}}$\
limit. This convergence suggests that in this specific estimation regime,
both Gaussian measurement schemes are highly effective at capturing the
available quantum information. In Fig. 20(c), the ratio $\tilde{R}$\ for
heterodyne measurement starts from a low initial value of approximately $0.4$%
.\textbf{\ }As $\epsilon $\ increases, the ratio exhibits a steep and
monotonic climb, quickly surpassing the homodyne measurement and
asymptotically approaching the theoretical maximum of unity. This behavior
confirms that heterodyne measurement is highly effective at saturating the
QCRB in the strong-amplification regime. Conversely, homodyne measurement
shows a non-monotonic trend with limited efficacy. The ratio initially
increases slightly, reaching a local peak of approximately $0.63$\ near $%
\epsilon \approx 0.4$, before entering a gradual decline and stabilizing at
a plateau of roughly $0.58$. The intersection of the two curves at small $%
\epsilon $\ highlights that while homodyne measurement may offer a slight
advantage in the weak-amplification limit, it is rapidly overtaken by
heterodyne measurement, which proves to be the superior strategy for
extracting quantum information as the gain of the thermal amplifier channel
increases.

Regarding the input squeezed vacuum state, we perform numerical evaluations
of the ratio $\tilde{R}$\ between the QCRB and the CRB changing with the
relevant physical parameters $r,$\ $\bar{N}_{\text{amp}},$\ and $\epsilon $,
as shown in Fig. 21. Our analysis indicates that the homodyne measurement is
incapable of simultaneously estimating $\epsilon $\ and $\bar{N}_{\text{amp}%
} $\ due to the singularity of the classical Fisher information matrix,
resulting in a divergent CRB. Consequently, we omit the homodyne measurement
results and restrict our discussion in Fig. 21 to the performance of
heterodyne measurement. In Fig. 21(a), the ratio $\tilde{R}$\ maintains a
remarkably high efficiency, with the ratio starting above $0.91$. As the
acceleration parameter $r$\ increases, the ratio exhibits a smooth,
monotonic rise. For $r$\ $>2$, the curve asymptotically approaches the
theoretical limit of unity, indicating that heterodyne measurement can
almost perfectly saturate the QCRB in the high-acceleration regime. In Fig.
21(b), the ratio $\tilde{R}$\ demonstrates a rapid and monotonic increase as 
$\bar{N}_{\text{amp}}$\ grows. The curve originates at a high initial value
of approximately $0.87$\ and quickly plateaus, asymptotically approaching
the unity limit for $\bar{N}_{\text{amp}}\geq 10$. In Fig. 21(c) the ratio $%
\tilde{R}$\ exhibits a non-monotonic behavior in the weak-amplification
regime. Starting from an initial value of approximately $0.96$, the curve
undergoes a slight dip, reaching a local minimum near $\epsilon \approx 0.5$%
. Beyond this point, the ratio demonstrates a steady, monotonic increase,
asymptotically approaching the theoretical maximum of unity as $\epsilon $\
grows further. Despite the minor fluctuation at small $\varepsilon $, the
ratio remains consistently high (above $0.94$) across the entire
investigated range. This indicates that for squeezed vacuum states,
heterodyne measurement provides an exceptionally efficient estimation
strategy that nearly saturates the QCRB, particularly as the gain of the
amplifier channel increases.

\section{Conclusions}

In summary, we investigate the influences of the Unruh effect on the
precision of parameter estimation within Gaussian channels, specifically
focusing on thermal attenuator and thermal amplifier configurations. For the
thermal attenuator channel, we first examine a single-parameter estimation
scenario involving the attenuation parameter. Numerical analysis
demonstrates that the Unruh effect detrimentally impacts estimation
precision; specifically, as the acceleration parameter increases, the QFI of
the attenuation parameter exhibits a monotonic numerical decline.
Furthermore, while homodyne measurement is more effective in regimes
characterized by low acceleration and small thermal mean numbers, the ratio
between the CFI for homodyne measurement and the QFI decreases as the
acceleration parameter and thermal mean number increase. In contrast, the
performance of heterodyne measurement improves with increasing acceleration
and thermal mean numbers, and it asymptotically saturates the QFI in the
limit of high acceleration parameter or high thermal mean numbers. We then
extend our investigation to a two-parameter estimation problem involving
both the attenuation parameter and the thermal mean number. Consistent with
the single-parameter case, the Unruh effect negatively impacts the joint
estimation precision, as evidenced by the numerical increase in the QCRB
with rising acceleration. For the thermal amplifier channel, we similarly
assess the impact of the Unruh effect on the QFI of the amplifier parameter,
finding that the QFI decreases gradually as the acceleration parameter
increases. In this context, heterodyne measurement consistently outperforms
homodyne detection and saturates the QFI in the limit of high acceleration,
high amplification, and large thermal mean numbers. Finally, we consider a
two-parameter estimation scenario involving the amplifier parameter and the
thermal mean number.\textbf{\ }The QCRB increases monotonically with the
acceleration parameter and approaches a finite, non-zero asymptotic value in
the limit of infinite acceleration. Our analysis further reveals a
progressive improvement in estimation precision as the amplification
parameter increases, and we confirm that heterodyne measurement saturates
the QCRB in the limit of high acceleration, high amplification, and large
thermal mean numbers.

Finally, we should mention that the results obtained in this paper can be
further interpreted within the broader context of relativistic quantum
information beyond the single-mode approximation. Several works have shown
that the complexities associated with mode structure can be addressed by
considering localized wave packets and localized detector-field interactions 
\cite{70,71,72}. These localized modes remain well defined for both inertial
and accelerated observers, ensuring a consistent physical interpretation of
the quantum states \cite{70,71,72}. Furthermore, it has been demonstrated
that the effect of relativistic acceleration on localized states can be
analytically derived as a Gaussian channel \cite{70,71,72}. Since the
transformation typically preserves the Gaussian character of the probe
states, our analysis of parameter estimation in thermal attenuator and
thermal amplifier channels remains highly relevant. Building upon these
findings, our future work will consider extending the present framework to
more complex multimode scenarios.

\begin{acknowledgments}
We sincerely thank Marco G. Genoni and Francesco Albarelli for their
discussions on the relevant work. This work was supported by the National
Key Research and Development Program of China (Grants No. 2021YFA1400900,
No. 2021YFA0718300, and No. 2021YFA1400243), NSFC (Grants No. 12074105, No.
12104135, No. 12404377, and No. 61835013), Natural Science Foundation of
Henan province (Grant No. 252300421995), and Key Scientific Re search
Projects of Henan Province Higher Education Institu tions (Grant No.
26B140007). Wei Ye is supported by Jiangxi Provincial Natural Science
Foundation (20232BAB211032, 20252BAC240169) and the Scientific Research
Startup \ Foundation (EA202204230).
\end{acknowledgments}

\textbf{Appendix\ A: Coefficients in the Eq.} (\ref{31})

In this Appendix, we provide the corresponding coefficients for the
commutation relation between symmetric logarithmic derivative operators $%
\hat{L}_{\vartheta }$ and $\hat{L}_{\bar{N}_{\text{att}}}$ in the Eq. (\ref%
{31}) 
\begin{align}
\Gamma _{1}& =\frac{\Lambda _{1}}{\Lambda _{4}^{2}\Lambda _{5}(1+\Lambda
_{5})^{2}},  \notag \\
\Gamma _{2}& =\frac{\Lambda _{6}}{\Lambda _{4}^{2}\Lambda _{5}(1+\Lambda
_{5})^{2}},  \notag \\
\Gamma _{3}& =\frac{\Lambda _{7}}{\Lambda _{4}^{2}\Lambda _{8}^{2}},  \notag
\\
\Gamma _{4}& =\frac{\Lambda _{9}}{\Lambda _{4}^{2}\Lambda _{8}^{2}}, 
\tag{A1}
\end{align}%
where

\begin{align}
\Lambda _{1}& =\sqrt{2}(\Lambda _{2}-\Lambda _{3})\text{Im(}\alpha \text{)}%
\cosh r\sin ^{3}\vartheta ,  \notag \\
\Lambda _{2}& =8-5\bar{N}_{\text{att}}^{2}+2\bar{N}_{\text{att}}^{2}\cosh
^{2}r[4\cos (2\vartheta )+\cos (4\vartheta )],  \notag \\
\Lambda _{3}& =\cosh (2r)[8+\bar{N}_{\text{att}}(16+5\bar{N}_{\text{att}})],
\notag \\
\Lambda _{4}& =2+\bar{N}_{\text{att}}-\bar{N}_{\text{att}}\cos (2\vartheta ),
\notag \\
\Lambda _{5}& =(\bar{N}_{\text{att}}\cos (2\vartheta )-\bar{N}_{\text{att}%
}-1)\cosh ^{2}r-\sinh ^{2}r,  \notag \\
\Lambda _{6}& =-\sqrt{2}(\Lambda _{2}-\Lambda _{3})\text{Re(}\alpha \text{)}%
\cosh r\sin ^{3}\vartheta ,  \notag \\
\Lambda _{7}& =-16\sqrt{2}\text{Im(}\alpha \text{)}\bar{N}_{\text{att}}\text{%
sech}^{3}r\cos ^{2}\vartheta \sin ^{3}\vartheta ,  \notag \\
\Lambda _{8}& =1+\bar{N}_{\text{att}}-\bar{N}_{\text{att}}\cos (2\vartheta )-%
\text{sech}^{2}r+\tanh ^{2}r,  \notag \\
\Lambda _{9}& =16\sqrt{2}\text{Re(}\alpha \text{)}\bar{N}_{\text{att}}\text{%
sech}^{3}r\cos ^{2}\vartheta \sin ^{3}\vartheta .  \tag{A2}
\end{align}

\textbf{Appendix\ B: Coefficients in the Eq.} (\ref{44})

In this Appendix, we present the detailed coefficients corresponding to the
commutation relation between the symmetric logarithmic derivative operators $%
\hat{L}_{\epsilon }$ and $\hat{L}_{\bar{N}_{\text{amp}}}$ in the Eq. (\ref%
{44})

\begin{align}
\Phi _{1}& =\frac{\Psi _{1}(\Psi _{4}-\Psi _{5})}{\Psi _{6}},  \notag \\
\Phi _{2}& =\frac{\Psi _{2}(\Psi _{4}-\Psi _{5})}{\Psi _{6}},  \notag \\
\Phi _{3}& =\frac{\Psi _{7}\Psi _{9}}{\Psi _{10}},  \notag \\
\Phi _{4}& =\frac{\Psi _{8}\Psi _{9}}{\Psi _{10}},  \tag{B1}
\end{align}%
where%
\begin{align}
\Psi _{1}& =-4\sqrt{2}\text{Im(}\alpha \text{)}\cosh ^{3}r\sinh ^{2}\epsilon
,  \notag \\
\Psi _{2}& =4\sqrt{2}\text{Re(}\alpha \text{)}\cosh ^{3}r\sinh ^{2}\epsilon ,
\notag \\
\Psi _{3}& =\sinh ^{2}r+\cosh ^{2}r[(1+\bar{N}_{\text{amp}})\cosh (2\epsilon
)-\bar{N}_{\text{amp}}],  \notag \\
\Psi _{4}& =\frac{\sinh \epsilon }{\Psi _{3}},  \notag \\
\Psi _{5}& =\frac{2(1+\bar{N}_{\text{amp}})\cosh ^{2}r\cosh \epsilon \sinh
(2\epsilon )}{\Psi _{3}^{2}-1},  \notag \\
\Psi _{6}& =\Psi _{3}^{2}-1,  \notag \\
\Psi _{7}& =-8\sqrt{2}\text{Im(}\alpha \text{)}(1+\bar{N}_{\text{amp}})\cosh
^{5}r,  \notag \\
\Psi _{8}& =8\sqrt{2}\text{Re(}\alpha \text{)}(1+\bar{N}_{\text{amp}})\cosh
^{5}r,  \notag \\
\Psi _{9}& =\cosh \epsilon \sinh ^{2}\epsilon \sinh (2\epsilon ),  \notag \\
\Psi _{10}& =(\Psi _{3}^{2}-1)^{2}.  \tag{B2}
\end{align}

\end{document}